\documentclass[aps,nofootinbib,pre,reprint,superscriptaddress,longbibliography]{revtex4-2}
\usepackage{amsmath}
\usepackage{graphicx}
\usepackage{amsfonts}
\usepackage{color}
\usepackage{bm}
\usepackage{lipsum}
\usepackage{comment}
\usepackage{dsfont}
\usepackage{bm}
\usepackage{amsmath}
\usepackage{amssymb}
\usepackage{graphicx} 
\usepackage{braket}
\usepackage{mathtools}
\usepackage{times}
\usepackage{enumerate}
\usepackage{subcaption}

\newcommand{\bea}{\begin{eqnarray}}
\newcommand{\eea}{\end{eqnarray}}
\usepackage[colorlinks,linkcolor=blue,urlcolor=blue,citecolor=blue]{hyperref}
\usepackage{multirow}
\begin{document}

%


\title{Design Principles for Enhanced Quantum Transport with Site-Dependent Noise}




\author{Maggie Lawrence}
\email{maggie.lawrence@mail.utoronto.ca}l
\affiliation{Department of Physics and Centre for Quantum Information and Quantum Control, University of Toronto, 60 Saint George St., Toronto, Ontario, M5S 1A7, Canada}
\affiliation{Vector Institute, W1140-108 College Street, Schwartz Reisman Innovation Campus Toronto, Ontario M5G 0C6, Canada}

\author{Elise Wang}
\email{elisew.wang@mail.utoronto.ca}
\affiliation{Department of Chemical and Physical Sciences, University of Toronto Mississauga, 3359 Mississauga Road, Mississauga, ON, L5L 1C6, Canada}

\author{Dvira Segal}
\email{dvira.segal@utoronto.ca}
\affiliation{Department of Chemistry, University of Toronto, 80 Saint George St., Toronto, Ontario, M5S 3H6, Canada}
\affiliation{Department of Physics and Centre for Quantum Information and Quantum Control, University of Toronto, 60 Saint George St., Toronto, Ontario, M5S 1A7, Canada}

\date{\today}

\begin{abstract}
{
Environmental noise can enhance transport, an effect known as environmental noise-assisted quantum transport. Most theoretical studies focus on optimizing system parameters under spatially uniform system--environment coupling. Here, instead, we optimize the environmental noise itself by allowing for site-dependent dephasing.
We investigate steady-state transport in one-dimensional lattices with either ramped or disordered energy landscapes, considering both short- and long-range coherent tunneling. In the absence of environmental effects, in the thermodynamic limit these systems can exhibit localization, and thus suppressed transport, 
arising from destructive interference. Using a Lindblad master equation framework, we implement local dephasing optimized to maximize steady-state population flux.
We find that for ramp potentials, short-range tunneling favors selective dephasing on alternating sites, whereas long-range tunneling benefits from a dephasing profile that increases with distance from the injection site. In energetically disordered systems, strongly detuned sites require enhanced local dephasing under short-range tunneling to facilitate transport. 
In all cases, we find that site-optimized dephasing allows higher transport efficiency than uniform dephasing, and it is accompanied by increased spatial delocalization of the steady state. Our results provide microscopic insight into the interplay between coherent dynamics and environmental noise. Dephasing broadens energy levels locally, helping to overcome detuning and destructive interference.
More generally, we establish spatially-structured environmental noise as a strategy for controlling both quantum transport and state coherence in open systems.
}
\end{abstract}
\maketitle

\section{Introduction}
Devices such as organic solar cells \cite{Creatore2013,Fruchtman2016,DeSio2017,Rouse2019,Cavassilas2020,Hu2021QDTransport} or quantum dots-based charge transport systems \cite{Lei2022ChargeTransportQD,Yang2015QDEfficiency,Bush2021,Bahadou2025,ContrerasPulido2017,Wang2007,Abdullah2016,Mathe2022,Chen2013,Eastham2013} rely on efficient particle transfer across networks governed by quantum dynamics.
Intuitively, one would think that limiting external effects such as noise would be the best way to promote transport in such devices; however,
quantum-coherent evolution is not always the optimal mechanism for efficient carrier transfer, particularly in regimes where quantum transport depends on deep tunneling processes or where quantum effects ``conspire" to produce destructive interference. In extended systems, a substantial body of work has shown that coherent quantum dynamics can inhibit transport through interference-induced localization. For example, in low dimensional systems, uncorrelated static disorder can lead to an exponential localization of single-particle wavefunctions, a phenomenon first predicted by Anderson \cite{anderson58,Kramer93,Lahini08andersonphotonic,Schwartz2007anderson}. In this case, destructive interference between multiple scattering pathways confines the particle to a finite spatial region.
A similar suppression of coherent transport arises in lattices subjected to a uniform potential gradient (ramp potential). In such systems, Bloch oscillations lead to the so-called Wannier-Stark (WS) localization \cite{Wannier62,emin87,Gluck02,Bandyopadhyay2021WS,Erik2020JCP, Erik2022PRXQ}, wherein the particle remains confined due to coherent phase winding induced by the energy tilt, rather than disorder.  Localization and suppression of transport were also examined in many other systems, emerging due to, e.g., incommensurate potential with the lattice, such as in the Aubry–André or Harper models \cite{AA80,Harper1955,Sarma17}, many body interactions \cite{Abanin19,MBLexp,Mirlin05}, and under time periodical driving \cite{Abanin16}.
Specifically for Anderson and Wannier-Stark models, localization effects were experimentally observed in a variety of physical systems, such as photonic lattices \cite{PL99,OL07}, waveguide arrays \cite{Martin2011opticsexpress,Gao2023WS,Mukherjee2015}, ultracold atoms \cite{Billy2008Anderson,chabe2008PRL, moore1994PRL, Manai2015PRL}, and more recently superconducting circuits \cite{SC22,Guo2021WS,Song2024PRXWS}.

In most applications of quantum systems, interactions between charge carriers and their surrounding environment are unavoidable. These interactions, commonly referred to as environmental noise, give rise to decoherence, dephasing, and energy dissipation. For processes that rely on quantum coherence and energy conservation, such effects are typically detrimental, limiting performance. 
%
However, contrary to this conventional view, many studies have revealed a counterintuitive phenomenon known as environmental-noise-assisted quantum transport (ENAQT). In the ENAQT regime, system–environment interactions enhance transport, defining an ideal operating region. This effect typically manifests itself as a turnover behavior in a transport figure of merit, such as flux, efficiency, or current, when analyzed as a function of the environmental coupling strength. Having been extensively discussed theoretically in the context of light-harvesting complexes \cite{Alan08,Plenio08,Plenio09,kassal09, Plenio10,CaoEET1, CaoEET2, CaoEET3,Plenio12,Montiel2014,Plenio21,mohseni2014energy,zerah-harush2018, manzano2013quantum, Kassal2012ordered, Erik2023PRXE, GaugerKassal2021JPCL}, ENAQT has since been experimentally observed on a variety of platforms, including photonic systems \cite{cavity15,biggerstaff2016waveguides,photon24}, trapped-ion networks \cite{maier2019environment}, nanocrystal superlattices \cite{Blach25}, and superconducting circuits \cite{sundelin2026quantumrefrigeration}. Indeed, dephasing from environmental fluctuations disrupts the phase coherence underlying the WS and Anderson localization, see, e.g., Refs. 
\cite{WSD23,GooldF24,WSD24,And10,And13,Coates2021NJP},
restoring extended transport where an optimal \textit{balance} between coherent interference and incoherent effects maximizes mobility.

Having this balance is key: Although weak coupling facilitates transport in low-dimensional disordered systems, excessively strong interactions lead to suppression, attributed to the quantum Zeno effect \cite{zerah-harush2020}. This characteristic turnover behavior has a long history in classical stochastic systems, where it is known as Kramers’ turnover \cite{kramers1940brownian,Pollak89,Haenggi90,NovotnyExp17}, as well as in quantum charge and energy transport systems \cite{Segal00,Cao13}; these studies represent a few examples of a large body of related work. In both Anderson and Wannier-Stark models, coherent localization results from the {\it maintenance of phase in the wavefunction}, and hence it can be disrupted by coupling the system to an environment. 
 
Typically, theoretical studies of ENAQT assume a uniform system–environment coupling and investigate which system structures yield enhanced transport at intermediate noise strengths \cite{Cao2009,Kassal2012ordered,manzano2013quantum,mohseni2014energy,manzano2016lattice,zerah-harush2018,zerah-harush2020,Plenio21,kurt2023, Shabani2014numerics,Coates2021NJP}. Recent work has shown that energy landscapes that optimize quantum transport under specific environmental conditions can be highly nontrivial \cite{ML25}. However, these studies still assume spatially uniform system–environment interactions. This raises a natural question: can transport be further enhanced through a richer coherent–incoherent mechanism that balances unitary dynamics with site-engineered environmental noise? For instance, might transport benefit from preserving coherence over several sites, rather than subjecting every site to identical environmental dephasing? 
This question has been addressed indirectly for the Fenna–Matthews–Olson (FMO)
photosynthetic complex \cite{Alan08, Plenio08, Plenio09, Plenio10, Plenio12}. Some studies have also suggested that local tuning of environmental noise may help explain long-range charge transport in biological systems \cite{Beratan19,Eshel20} such as cable bacteria \cite{Naggar24}. 
Despite extensive studies of ENAQT, it remains unclear whether and how spatially structured environmental dephasing can be engineered to overcome localization and enhance transport.


In this work, we explore local engineering of environmental noise to enhance quantum transport. Specifically, we consider one-dimensional chains with Coulomb- or dipole-like tunneling between sites and optimize the environmental dephasing of energy levels at each site to maximize the transport flux. Our goal is to provide
an understanding of the design principles underlying optimal noise profiles. 
Environmental effects are modeled using a Lindblad quantum master equation. We consider two types of system that show localization in one dimension in the thermodynamic limit when subjected to coherent short-range tunneling: Chains with a ramp-like energy landscape, showing Wannier-Stark localization, and energy-disordered chains, leading to Anderson localization. Comparing the steady state flux under either site-optimized dephasing or uniform dephasing, we find that site-optimized dephasing rates provide physical insights and guidelines on optimizing transport. In particular, we find that locally-optimized dephasing profiles consistently create steady states with more extended coherences than the case of uniform dephasing and can lead to significant increases in carrier flux.
Our main conclusion as such is that structured environmental noise is a viable control knob for tuning both transport efficiency and quantum coherence.

The structure of this paper is as follows.
In Sec. \ref{sec:Model}, we present the model, the equation of motion, the optimization approach and the observables. In Sec. \ref{sec:WS}, we focus on overcoming the Wannier-Stark localization by site-dependent optimized dephasing. We begin with an analytically solvable three-site model and continue with numerical simulations. In Sec. \ref{sec:Anderson} we consider energy-disordered chains that demonstrate Anderson localization in one dimension in the quantum coherent limit. We first present an example and display its site-engineered design. Then, we draw general conclusions from ensemble properties: Averaged coherence length and the correlation of locally enacted dephasing with local energy mismatches. We summarize our findings in Sec. \ref{sec:Summ} offering directions for future studies. 

\section{Model, Method and observables}
\label{sec:Model}

\subsection{Quantum Transport model}

We simulate carrier (charge, excitation) transport in quasi-one-dimensional systems in which each site is coupled to an independent local environment. Our goal is to identify site-dependent environmental noise levels, modeled as dephasing, which enhance particle transfer from one end of the chain to the other. We refer to the model as quasi-one-dimensional since we consider increasingly long range tunneling. Under the most extended connectivity, the system effectively develops an all-to-all connectivity between sites.

We introduce several simplifying assumptions that are relevant for charge and exciton transport and make the problem computationally manageable. First, we allow only for a single particle within the whole quantum system, corresponding to a single excitation or a single charge. This is a physically reasonable assumption; for example, in the Fenna-Matthews-Olson protein complex, which was extensively studied in the context of photosynthetic function, the time it takes an exciton to propagate from the antenna to the reaction center is much shorter (subpicoseconds to picoseconds) than the time at which photons are absorbed in the protein under physiological conditions (microseconds or longer). This implies that energy transfer is typically completed before the arrival of the next excitation \cite{Scholes14}. 
Second, we model the interactions between the quantum system and its environment using the Lindblad quantum master equation \cite{breuer2007_decoherence}. This description applies to different environments, such as electrons \cite{Rosati2014}, photons \cite{giri2025entanglement}, phonons \cite{Jager2022}, and intra- or inter-molecular vibrations \cite{sarkar2020environment, juhasz2018vibrations}. In what follows, we work with units of \(\hbar \equiv 1\).

In the $N$-site chain's site-local basis, the Hamiltonian of the system is given by a tight-binding model,
\begin{equation}
    \hat{H}_S = \sum_{n=1}^N \varepsilon_n \ket{n} \bra{n} + \sum_{n\neq m} J_{|n-m|} \ket{n} \bra{m}.
\end{equation}
Here, \(\varepsilon_n\) is the on-site energy of site \(n\), and \(J_{|n-m|}\) is the tunneling strength between sites \(m\) and \(n\). Below we consider two models for the energy profile.  In the Wannier-Stark model, we build a ramp potential using $\varepsilon_{n}-\varepsilon_{n+1}=\Delta$, with $\Delta >0$, and we set as an energy scale $N\Delta = 1$ (such that the total gap is $(N-1)\Delta$).  
In the Anderson model, we sample energy levels from a uniform distribution, $\varepsilon_n \sim \mathcal{U}(0,1)$, independently for all $n$.

For simplicity, we choose these tunneling elements to be real and positive. To mimic real systems with, e.g., dipolar interactions,  we model these tunneling energies using a power-law function,
\begin{equation}
    J_{|n-m|} = \frac{J_{max}}{|n-m|^\alpha}.
    \label{eq:tunnel}
\end{equation}
Under this model, the maximum tunnel strength \(J_{max}\) is observed between nearest neighbors. We characterize our systems as long-range or short-range based on the power in the denominator: \(\alpha = 1\) leads to long-range couplings, \(\alpha = 5\) represents short-range coupling, and $\alpha=3$ an in-between value.

To model system-environment interactions (also referred to as ``noise"), we use local dephasing effects, as enacted, e.g., in trapped-ion experiments \cite{maier2019environment}. We assume that the environment acts locally and independently on each site to dephase the state, such that the local Lindblad QME is given by 
\begin{equation}
    \dot{\rho} = -i \left[ \hat{H}_S, \rho\right] + \sum_{n=1}^N \Gamma_n \left( \hat{L}_n \rho \hat{L}^\dagger_n - \frac{1}{2} \left\{\hat{L}^\dagger_n \hat{L}_n, \rho \right\} \right).
    \label{eq:lindblad}
\end{equation}
Here, \(\hat{L}_n = \ket{n} \bra{n} \) are site-local Lindblad jump operators, with corresponding dephasing rate constants \(\Gamma_n\). The dynamics evolve in the site basis; in the energy basis, this local dephasing model corresponds to an unstructured infinite-temperature bath. With no other mechanisms to add or remove particles, the steady-state solution of the dynamics corresponds to the maximally-mixed state, regardless of the system Hamiltonian. 
To describe a steady state transport process, we further introduce carrier injection and extraction Lindblad operators into our model,  corresponding to a trapping and renewal process:
$\hat{L}_l = \sqrt{\gamma_l}\ket{1}\bra{N}$.
%
This operator represents a jump between the two ``ends" of the chain, from site \(N\) to site 1. It enforces a constant flow of carriers entering site 1 and leaving site \(N\), both at a rate constant \(\gamma_l\). Different combinations of injection and extraction sites may lead to different transport behaviors; see Refs. \cite{maier2019environment, Dutta2021OutOfEquilibrium}. We chose the ends of the chain for the extraction and injection sites, in order to achieve transport over the longest distance allowed. Typically, $\gamma_l$ is chosen in a regime where it is not too small to become the rate determining step, thus allowing the structure of the internal system and the energy parameters to dictate the efficiency of the transport. From the other end, very large $\gamma_l$ would lead to energy broadening, masking the details of the energy profile and the dephasing effect.
With this additional trapping-renewal process, with a rate constant $\gamma_l$, the steady state deviates from the completely mixed state and it depends on the magnitude of $\gamma_l$ \cite{Schaller22,ML25}.

Our figure of merit for transport is the flux of population out of the \(N^\textrm{th}\) site of the system, 
\begin{equation}
    \eta = \gamma_l \rho_{NN}^{{\rm SS}},
    \label{eq:eta}
\end{equation}
where \(\rho_{NN}^{{\rm SS}}\) is the steady-state population of site \(N\), calculated by solving Eq. (\ref{eq:lindblad}) in the steady state. To simplify the notation, and since we only consider steady-state scenarios, below we use $\rho$ to refer to the steady-state solution.

The goal of this work is to maximize this quantity as a function of dephasing rate constants \(\Gamma_n\), which could be distinct at each site. By examining the resulting dephasing profiles and the properties of the  steady state, we aim to reveal whether and how environmental noise can be exploited to control for quantum transport and coherence.

Regarding energy scale for parameters:
For the ramped setup, we set the energy gaps in the $N$-site chain to
$\Delta = 1/N$, leading to a total energy bias $\sim 1$. Similarly, for energy-disordered systems, we sample energies from a uniform distribution $\mathcal U(0,1)$, which again sets the relevant energy scale. 
Other parameters are set relative to these energy scales, and we work in a regime where nearest neighbor tunneling energies are of the same order of magnitude as local energy gaps, facilitating coherent transport.
Our results further show that uniformly-optimized or site-optimized dephasing rates are of the same order of magnitude as these other energy parameters (energy gaps and tunneling energies), leading to ENAQT dynamics benefiting from both coherent and incoherent effects.

\subsection{Optimization approach} \label{subsec:opt}
The optimal maximum population flux is obtained using the Adamax iterative optimization algorithm. The derivatives of \(\eta\) are taken with the JAX library, and the optimizer was implemented with the JAX Optax library \cite{deepmind2020jax}. Since environmental noise is typically analyzed on a logarithmic scale, we performed our gradient ascent with respect to \(\log_{10}\Gamma\).

The high dimensional flux function $\eta$, depending on all local dephasing rate constants, may have many local maxima. To limit our solutions to the realm of ``physical" possibilities, we place hard bounds on all \(\Gamma_n\) to be between \(10^{-7}\) and 1. For the ramp-like system, we run the optimization algorithm 100 times beginning with different initial guesses for the optimum, and choose the best solution found by the optimizer. For the disordered system, we take 500 independently generated energy profiles, find the single uniform dephasing \(\Gamma_\mathrm{u}\) that maximizes population flux first for each of them, then we optimize the population flux as a function of all local dephasing rate constants \(\Gamma_{n}\) beginning from this optimal \(\Gamma_\mathrm{u}\). 

The solutions may include the boundary of our acceptable range for \(\Gamma_n\), i.e. \(\Gamma_n = 1\) or \(10^{-7}\). Therefore, we may not be reaching the global maximum but one of the local-maxima or boundary solutions. We set a minimum number of 30 steps for the optimizer, up to a maximum of 100 000. Any single \(\Gamma_n\) reaching a boundary value, the optimizer reaching 100 000 steps, or all gradients of \(\eta\) with respect to \(\Gamma_n\) being less than \(10^{-8}\) were stopping conditions for the optimizer. We considered an optimization ``successful" if the latter condition is met. In practice, most ramped system solutions converged to the same maximum, which we take to be the global maximum, for each tunneling coupling scheme considered. Most disordered systems had at least one \(\Gamma_n\) reaching the upper boundary value of 1, and therefore did not find a true ``maximum"; however, all of these solutions that included the optimizer boundary were still more efficient for transport than when each site was subject to the same \(\Gamma_\mathrm{u}\).


\section{Enhancing Transport In the Ramped System}
\label{sec:WS}

Consider a one-dimensional lattice subject to a uniform potential gradient, which results in a ramp-like energy profile. In the regime of fully quantum-coherent transport, in the thermodynamic limit particles will become spatially localized in this system, a phenomenon known as Wannier–Stark localization \cite{Wannier62,emin87,Gluck02}.
Because this localization arises from destructive interference, it can, in principle, be mitigated by dephasing. 

Our goal in this section is to identify site-dependent dephasing regimes that maximize exciton flux in these ramped systems.
We begin by presenting an approximate analytical solution for transport in a three-site model. The insights gained from this minimal system are then used to interpret numerical results for twelve-site chains. Throughout, we compare long-range and short-range tunneling scenarios, revealing distinct transport mechanisms and optimization strategies.

\subsection{Lessons from the Three-Site System}
\label{sec:3sites}

\begin{figure*}[htbp]
\centering
    \includegraphics[width=1\linewidth] 
{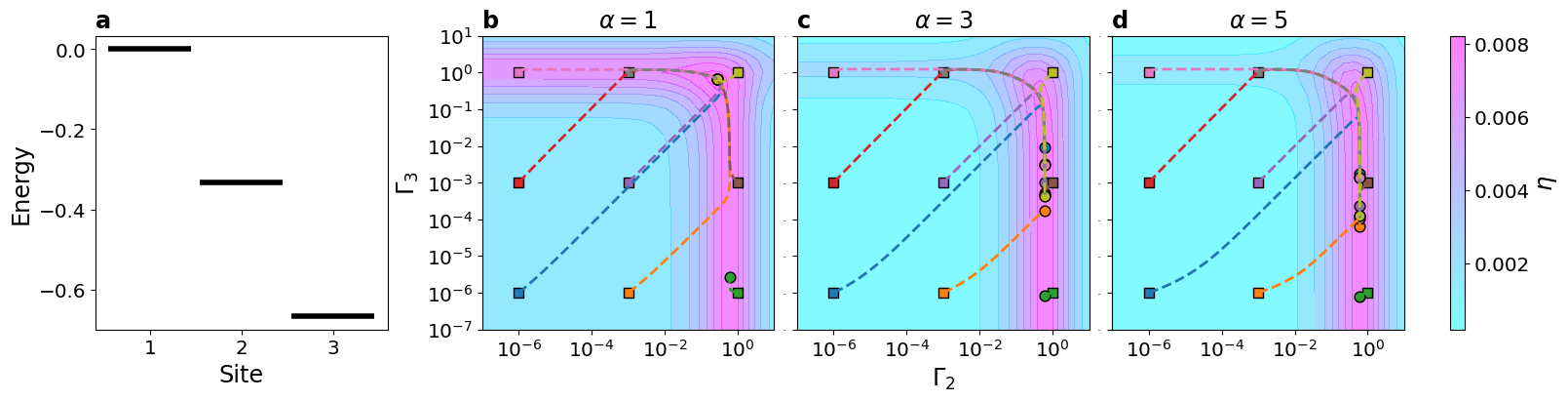}
\caption{Three-site example of transfer behavior. (a) Energy structure in the form of a ramp with equidistant energy levels. 
(b)-(d) Population flux as a function of dephasing rate applied onto sites 2 and 3. No dephasing is applied to site 1. Between the different panels, we vary the power $\alpha$ that controls the tunneling range, see Eq. \eqref{eq:tunnel}: $\alpha=1$ corresponds to long range tunneling, while $\alpha=5$ essentially describes nearest neighbors tunneling. Dashed lines show example trajectories taken by the Adamax optimizer from a grid of initial $\Gamma_2$ and $\Gamma_3$ values (squares of different colors) and the optimal solutions reached (circles).  Parameters are $\gamma_l=0.1$ and  $J_{max} = 0.1$, fixed throughout this study. } 
\label{fig:figure1}
\end{figure*}

A three-level model under ramp potential is depicted in Fig. \ref{fig:figure1}(a). 
In the short-range tunneling case 
the model essentially reduces to the nearest neighbor tight binding model. In contrast, under long range tunneling, site 1 is also well coupled to site 3. We implement dephasing at sites 2 and 3 with strength $\Gamma_{2,3}$; for simplicity, we set $\Gamma_1=0$. 

The three-site model is simple enough to be solved analytically in some limits. 
Furthermore, it is used to  test the optimization protocol.


\subsubsection{Simulations}

In Figs.~\ref{fig:figure1}(b)-(d), we present the population flux $\eta=\gamma_l \rho_{3,3}$ as a function of $\Gamma_2$ and $\Gamma_3$; with only two free parameters, the population flux landscape can be easily visualized. The results are shown for three values of the tunneling range parameter, $\alpha = 1$, $\alpha = 3$, and $\alpha = 5$, corresponding to long-, intermediate-, and short-range tunneling,
respectively.
We also display examples of optimization ``trajectories".  As seen in Fig. \ref{fig:figure1}(b)-(d), some initial conditions converge to local optima rather than the global maximum.

In the regime of low environmental noise, coherent tunneling produces small flux, as evident in the bottom-left region of Fig. \ref{fig:figure1}(b)-(d); the large energy gap between the first and last site hinders transport, which in longer chains shows as particle localization. 

By scanning the individual dephasing rates $\Gamma_2$ and $\Gamma_3$ using the Lindblad quantum master equation, we observe significant differences between long-range ($\alpha=1$) and short-range ($\alpha=5$) tunneling scenarios:
For quasi-nearest-neighbor tunneling ($\alpha=5$), optimal transport occurs when dephasing is applied primarily to the middle site, while the exit site remains essentially noise-free, see Fig. \ref{fig:figure1}(d). In contrast, for long-range tunneling ($\alpha=1$), both the middle and exit sites benefit from dephasing, see Fig. \ref{fig:figure1}(b). This can be understood as follows: Under short-range tunneling, transport is maximized by broadening the middle level, allowing it to overlap with its neighbors, thus efficiently accepting population from site 1 and transferring it to site 3. Local dephasing acts as a level-broadening mechanism. In contrast, for long-range tunneling, excitations can be exchanged directly between sites 1–2 and 1–3. As such, the level-broadening of both intermediate and exit sites facilitates transport.
Another observation is that a broad range of dephasing magnitudes produces similar fluxes. 

These trends suggest general principles for enhancing transport in longer systems: (i) Under short-range tunneling, ``middle" levels are broadened to facilitate nearest-neighbor transfer. (ii) In long-range tunneling, enhanced dephasing across all levels promotes transport, as one can ``jump over" neighbors.

\subsubsection{Analytical results: Nearest neighbor tunneling}

We support numerical observations by analytically solving the approximate transport problem. First, we focus on the nearest neighbors (NN) case, corresponding to $\alpha=5$. Relevant simulation results are shown in Fig. \ref{fig:figure1}(d).
We use the energy structure as depicted in Fig. \ref{fig:figure1}(a), defining $\Delta = |\varepsilon_2-\varepsilon_1| =|\varepsilon_3-\varepsilon_2| $ and taking $J_2=0$. For simplicity, we set $\Gamma_1=0$.
We also assume that the tunneling energy and the leak rates are small compared to the energy level difference and the dephasing, $J_1, \gamma_l <|\Delta|, \Gamma_{2,3}$.
Solving the problem in steady state, we obtain the population of site 3, which allows us to calculate the population flux.
In second order in tunneling energy $J_1$, we get
\bea
\eta_{{\rm NN}} &= &
J_{1}^{2}
\frac{4 \Gamma_{2}\,(\Gamma_{2}+\Gamma_{3})}
{4 \Delta^2\,(3 \Gamma_{2}+\Gamma_{3}) + \Gamma_{2}\,\bigl(3 \Gamma_{2}^{2}+5 \Gamma_{2} \Gamma_{3}+2 \Gamma_{3}^{2}\bigr)
} 
\nonumber\\
&+&  O(J_1^3).
\label{eq:v1}
\eea
We simplify this expression in two limits. First, when $\Gamma_3$ is small, we get 
\bea
\eta_{\rm NN}(\Gamma_3<\Gamma_2)&=& J_{1}^{2}
\frac{4 \Gamma_{2}}{3 \left(4 \Delta^{2} + \Gamma_{2}^{2}\right)}
\nonumber\\
&-&
 J_{1}^{2}\frac{8 \left(-4 \Delta^{2} + \Gamma_{2}^{2}\right) \Gamma_{3}}
{9 \left(4 \Delta^{2} + \Gamma_{2}^{2}\right)^{2}} + O(\Gamma_3^2).
\label{eq:G3zero}
\eea
Depending on the competition between energy splitting ($\Delta$) and dephasing on the central site ($\Gamma_2$), the second term can be positive (enhancing flux when turning on $\Gamma_3$) or negative (reducing flux with increasing $\Gamma_3$). 
If $\Gamma_2>2\Delta$, the second term becomes negative. In this regime of large dephasing of the middle level, turning on $\Gamma_3$ hinders transport.

If we return to Eq. (\ref{eq:v1}) and enforce $\Gamma_2=\Gamma_3$ we get
\bea
\eta_{\rm NN}(\Gamma_2=\Gamma_3)=
\frac{4 \Gamma_{2,3}J_{1}^{2} }{8 \Delta^{2} + 5 \Gamma_2^{2}}.
\label{eq:G2G3}
\eea
Focusing on the limit $\Gamma_{2,3}>\Delta$, it is easy to verify from Eqs. (\ref{eq:G3zero}) and (\ref{eq:G2G3}) that $\eta_{\rm NN}(\Gamma_2=\Gamma_3)< \eta_{\rm NN}(\Gamma_3<\Gamma_2)$.

We conclude that for nearest neighbor tunneling, transport is more efficient when dephasing is enacted only on the middle level 2 while level 3 remains free of environmental noise \((\Gamma_3 < \Gamma_2)\), compared to the case when dephasing is active on both sites \((\Gamma_3 = \Gamma_2)\).

\subsubsection{Analytic results: Long range tunneling}

We now consider the case of long range (LR) tunneling, related to $\alpha=1$, with relevant results depicted in Fig. \ref{fig:figure1}(b). 
For simplicity, here we use identical values for the tunneling energies, $J=J_2=J_1$.
Under assumptions of small tunneling energy and leak rates compared to energy levels difference and dephasing, $J, \gamma_l <|\Delta|, \Gamma_{2,3}$, and,  to second order in the tunneling energy, we get
\begin{widetext}
\bea
\eta_{\rm LR}=
\frac{
8 \Bigl[\Gamma_{2}\Gamma_{3}\,(\Gamma_{2}+\Gamma_{3})^{2}
+ 2 \Delta^{2}\,(4 \Gamma_{2}^{2}+6 \Gamma_{2}\Gamma_{3}+\Gamma_{3}^{2})\Bigr]\, J^{2}
}{
\left(16 \Delta^{2}+\Gamma_{3}^{2}\right)
\left[4 \Delta^{2}\,(3 \Gamma_{2}+\Gamma_{3})
+ \Gamma_{2}\,(3 \Gamma_{2}^{2}+5 \Gamma_{2}\Gamma_{3}+2 \Gamma_{3}^{2})\right]
}
+O(J^3).
\label{eq:v1v2}
\eea
Expanding this result in $\Gamma_3$ we get
\bea
\eta_{\rm LR}(\Gamma_3<\Gamma_2)=
\frac{4 \Gamma_{2}\, J^{2}}{3 \left(4 \Delta^{2} + \Gamma_{2}^{2}\right)}
\;+\;
\frac{\left(112 \Delta^{4} + 8 \Delta^{2} \Gamma_{2}^{2} + 3 \Gamma_{2}^{4}\right)
J^{2}\, \Gamma_{3}}
{18 \Delta^{2} \left(4 \Delta^{2} + \Gamma_{2}^{2}\right)^{2}} + O(\Gamma_3^2).
\eea
\end{widetext}
Since the second term is positive, we conclude that implementing $\Gamma_3$ benefits transport.
Specifically, when $\Delta<\Gamma_2$
we get 
\bea
\eta_{LR}(\Gamma_3<\Gamma_2,\Delta < \Gamma_2)\approx 4J^2/3\Gamma_2.
\label{eq:v1v2g3zero}
\eea
For identical dephasing we get from Eq. (\ref{eq:v1v2}),
\bea
\eta_{\rm LR}(\Gamma_2=\Gamma_3)=
\frac{8 \Gamma_{2} \left(11 \Delta^{2} + 2 \Gamma_{2}^{2}\right) J^{2}}
{\left(16 \Delta^{2} + \Gamma_{2}^{2}\right) \left(8 \Delta^{2} + 5 \Gamma_{2}^{2}\right)},
\eea
thus, when, $\Delta<\Gamma_2$
\bea
\eta_{\rm LR}(\Gamma_3=\Gamma_2,\Delta<\Gamma_2)\approx 16J^2/5\Gamma_2,
\label{eq:v1v1g2g3}
\eea
which is larger than the comparable limit, Eq. (\ref{eq:v1v2g3zero}).
We conclude that under a long range tunneling process, particle transfer is higher when dephasing is enacted on both sites, 2 and 3, contrasting the nearest neighbor tunneling limit.

Generalizing, the main insight from this exercise is that in NN cases, dephasing need not be enacted at every site, as it is more beneficial to add it to ``middle" levels with broadening, thus allowing energy overlap with both back and forward neighbors. In contrast, LR models benefit from dephasing on every site, as this promotes transfer to distant sites. We now show that these guidelines are supported by numerical simulations on extended systems.
Our observations for the Wannier Stark-type ramp energy profile, under both short range and long range tunneling, are summarized in Table \ref{tab:3levels}.
\begin{table}[h]
\begin{tabular}{|c|c|}
\hline
Model                                       & Optimal dephasing regime \\ \hline
Nearest-neighbor tunneling, \(J_1 \gg J_2\) & \(\Gamma_3 < \Gamma_2\)  \\ \hline
Long-range tunneling, \(J_1 = J_2\)         & \(\Gamma_3 = \Gamma_2\)  \\ \hline
\end{tabular}
\caption{Summary of three-level model analytical results.}
\label{tab:3levels}
\end{table}


\subsection{Long Chains: Optimization of site-dependent dephasing }

\begin{figure*}[htbp]
    \centering
\includegraphics[width=\linewidth]{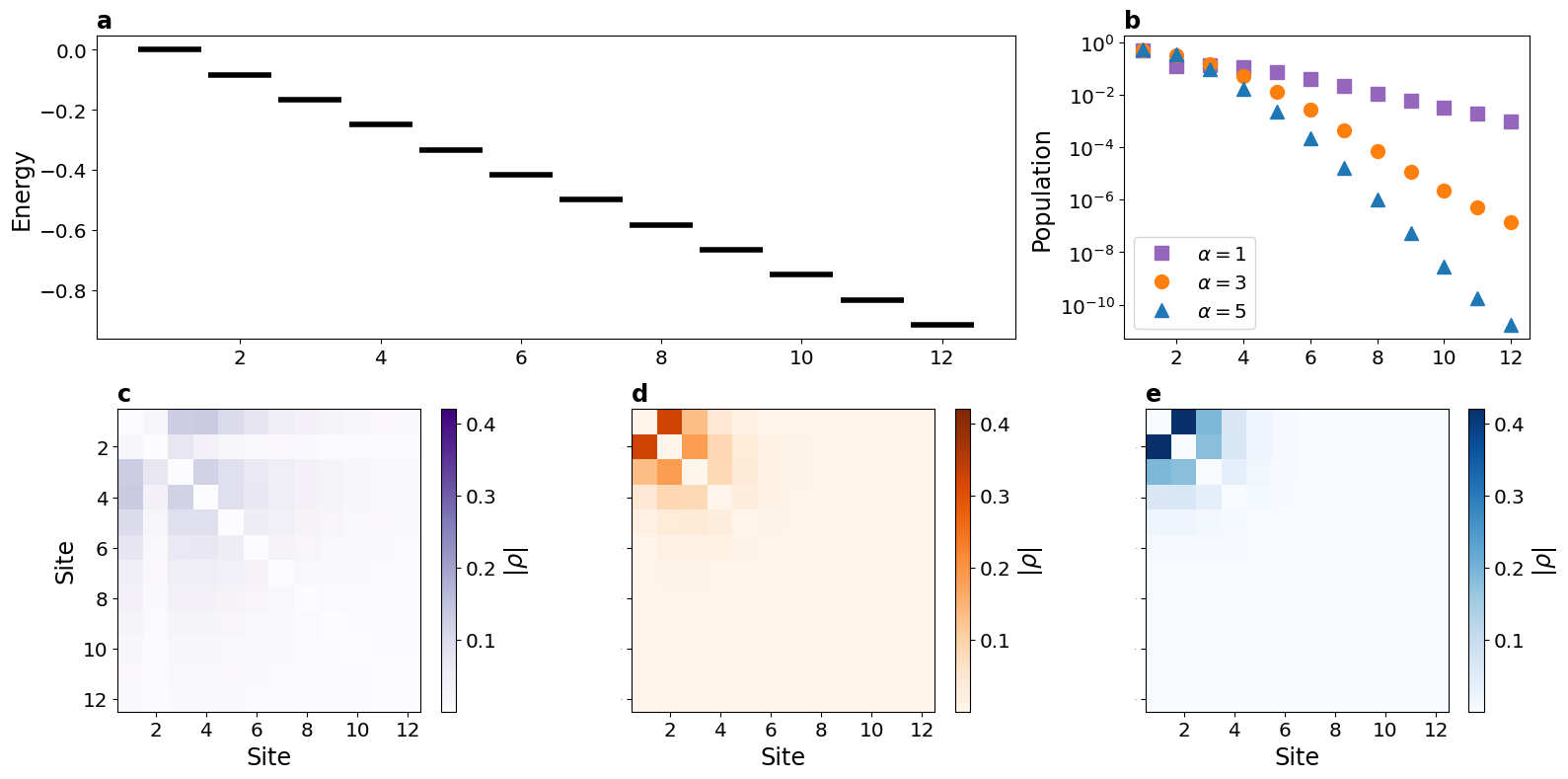}
    \caption{Development of WS localization: Lattices with $N=12$ sites and tunneling powers $\alpha = 1,3,5$, with no environmental interaction. (a) Potential profile, (b) steady state population profiles for the ramp energy potential, and (c)-(e) steady-state density matrices for tunneling powers of \(\alpha = 1, 3, 5\), respectively, with the diagonals removed. The system includes 12 sites, with $J_{max} = 0.1$ and $\gamma_l = 0.1$.
     }
\label{fig:figure2}
\end{figure*}

\begin{figure}[htbp]
    \centering   \includegraphics[width=0.7\linewidth] {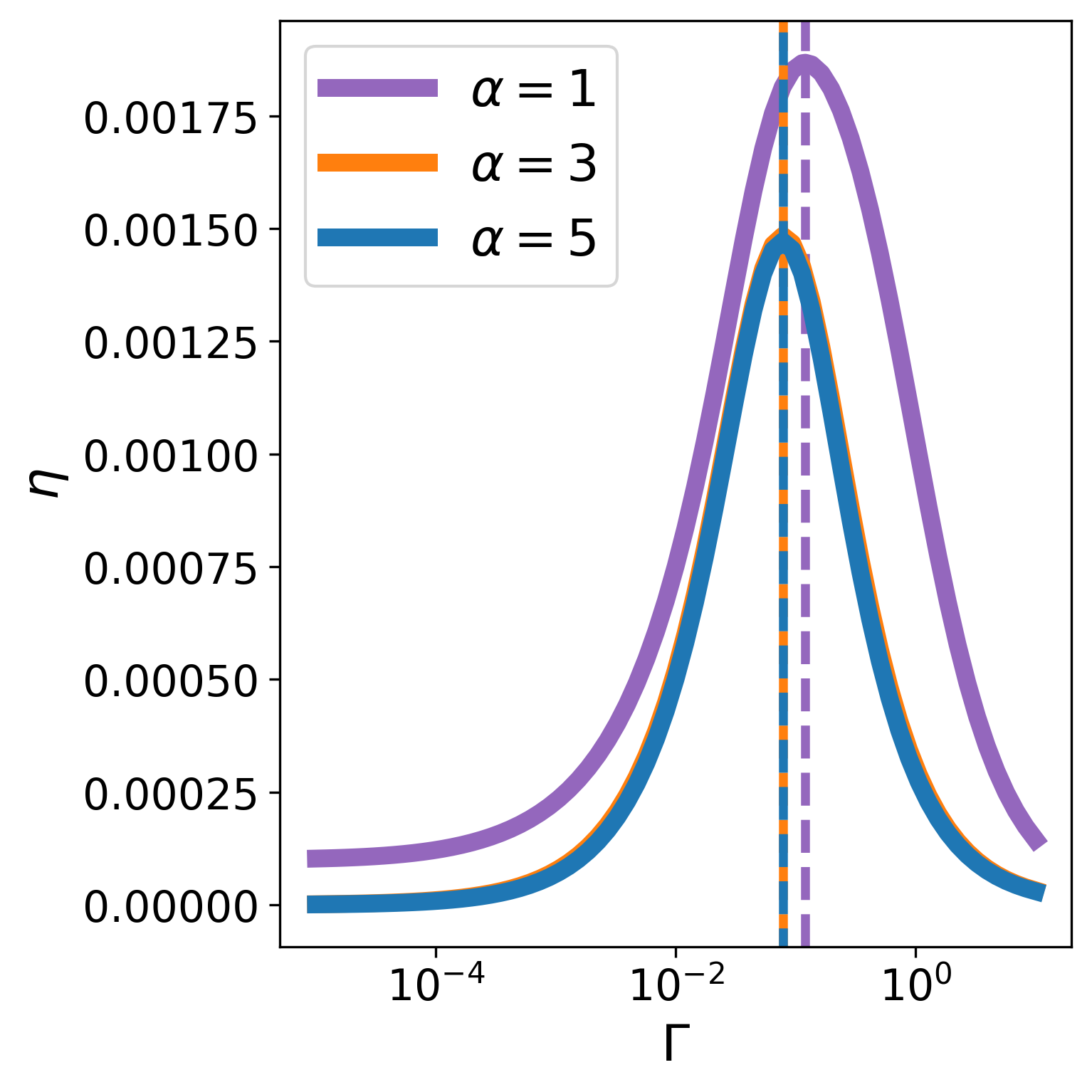} 
    \caption{
Population flux as a function of uniform dephasing under the ramp potential \ref{fig:figure2}(a). The peak indicates the optimal dephasing rate for transport. The dashed lines mark the locations of the peaks: \(\Gamma_{\rm u}^{\alpha=1} = 0.121\), \(\Gamma_{\rm u}^{\alpha=3} = 0.079\), and \(\Gamma_{\rm u}^{\alpha=5} = 0.078\) with corresponding fluxes $\eta_{\mathrm{u}}^{\alpha=1} = 1.87 \cdot 10^{-3}$, $\eta_{\mathrm{u}}^{\alpha=3}=1.48 \cdot 10^{-3}$, and $\eta_{\mathrm{u}}^{\alpha=5}=1.47 \cdot 10^{-3}$. Other parameters are the same as in Figure \ref{fig:figure2}.   
} 
    \label{fig:figure3}
\end{figure}



In this section, we examine the spatial profile of dephasing constants that can enhance transport in a system that would otherwise exhibit Wannier–Stark localization in the coherent limit. We restrict our analysis to finite-sized systems, specifically a chain of 12 sites. Such a configuration may serve as a model for a molecule \cite{Plenio08}, a short polymer \cite{Rouse2019}, or a chain of trapped ions \cite{maier2019environment}.


\begin{figure*}[htbp]
    \centering
\includegraphics[height=0.7\textheight]{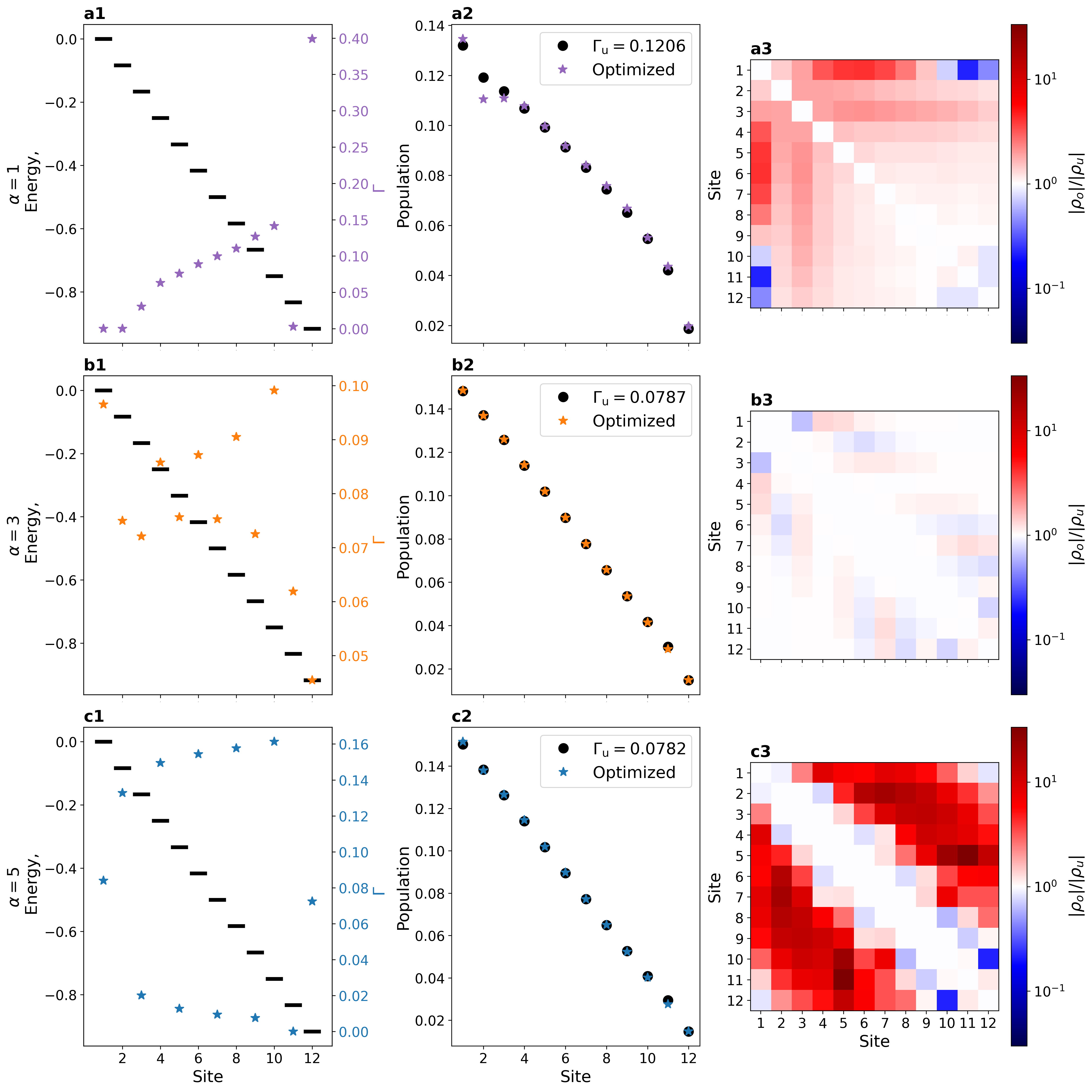} 
\caption{Enhancing transport in a ramp potential for systems with power-law tunneling  \(\alpha = 1\) (a1-a3, purple), \(\alpha = 3\) (b1-b3, orange), and \(\alpha = 5\) (c1-c3, blue).
Left column: We show the ramp energy profile and the resulting locally-optimized values for $\Gamma_n$.
Middle column: Steady state population using the optimal uniform value $\Gamma_\mathrm{u}$ and the optimized profile $\Gamma_n$.
Right column: Ratios of the absolute values of the density matrix elements
under the locally optimized dephasing rates, $\rho_o$, and the uniform value $\Gamma_\mathrm{u}$, $\rho_u$.
Other parameters are the same as in Figure \ref{fig:figure2}. We report here the values of the population flux $\eta$ under optimized and uniform dephasing profiles (optimized, uniform):    
$\eta^{\alpha = 1}: (1.97 \cdot 10^{-3}, 1.87 \cdot 10^{-3})$;
$\eta^{\alpha = 3}: (1.49 \cdot 10^{-3}, 1.48 \cdot 10^{-3})$;
$\eta^{\alpha = 5}: (1.49 \cdot 10^{-3}, 1.47 \cdot 10^{-3})$.
} 
\label{fig:figure4}
\end{figure*}
We begin by examining the emergence of localization in the absence of dephasing for chains with power-law tunneling characterized by exponents $\alpha = 1, 3, 5$. Figure~\ref{fig:figure2} shows (a) the energy profile, (b) the steady-state population profile, and (c)-(e) the corresponding maps of steady-state coherences. In the short-range case ($\alpha = 5$, Fig. \ref{fig:figure2}(e)), the population decays exponentially with distance from the injection site (site 1), and the coherences are tightly localized near the exciton origin. For intermediate-range coupling ($\alpha = 3$, Fig. \ref{fig:figure2}(d)), both the population and coherences decay more gradually with distance from the injection site, although the state remains localized. In the long-range regime ($\alpha = 1$, Fig. \ref{fig:figure2}(c)), the state develops pronounced tails extending across the chain, while the population remains peaked near the injection site. 
Note that our objective here is not to strictly identify localized or conducting phases, but to generally enhance transport. We are not focusing on the thermodynamical limit, but finite size systems, which we can readily optimize numerically, and are relevant for experiments. 

We now apply {\it uniform} dephasing on the 12-site system. In Fig.~\ref{fig:figure3}, we present the population flux as a function of the uniform dephasing constant \(\Gamma\). As before, we consider three values of the tunneling range parameter $\alpha$. In all cases, we observe a characteristic ENAQT behavior: at weak dephasing, the flux is suppressed due to localization. At strong dephasing, transport is again reduced, a manifestation of the quantum Zeno effect. Optimal transport occurs when $\Gamma$ is comparable to the level energy spacing $\Delta$, here around $\Gamma \sim 0.1$. We refer to the optimal uniform value as $\Gamma_{\rm u}$.

The question we now address is whether site-dependent dephasing rates, $\Gamma_n$, can further enhance transport performance, and, if so, what transport strategies emerge.
To this end, we introduce dephasing at each site and optimize the local values $\Gamma_n$ to maximize the population flux; results of this 12-parameter optimization are presented in Fig.~\ref{fig:figure4}. This solution was consistently achieved using 100 randomized trials.  

We first consider the long-range tunneling case, shown in Fig.~\ref{fig:figure4}(a1)-(a3). Here, we find that the flux is optimized when the dephasing strength increases mostly monotonically with the distance from the injection site (with the exception of the boundary site). Specifically, the dephasing begins near $\Gamma_1 \approx 0$, gradually increases to $\Gamma_{10} \sim 0.15$, then rises more sharply before dropping at the boundary, $\Gamma_{12} \approx 0$. 
This strategy can be rationalized noting that the system can transport coherently using long range connections. Thus, it is necessary to broaden more remote sites, which are too far off resonance for effective transport. The best strategy is then to increase dephasing (energy broadening) monotonically over the lattice. As such, the system benefits from long range tunneling in the vicinity of the injection site, yet it combats localization as sites become too far detuned. 

The corresponding steady state population profile is shown in panel (a2), where it is compared to the case of uniformly optimized dephasing  (the value leading to the peak in Fig. \ref{fig:figure3}). Although the site-optimized profile exhibits a slight improvement in population flux, the improvement over the uniform case is marginal, on the order of $\sim 5\%$.
The coherence map in panel (a3) shows the matrix elements $|(\rho_o)_{m,n}| / |(\rho_u)_{m,n}|$, where $\rho_o$ denotes the state under locally optimized dephasing and $\rho_u$ the state under uniformly optimized dephasing. The predominantly ``red'' character of the map indicates that the optimized solution enhances the extended coherences relative to the uniform case. The absolute values of coherences in the site-optimized case are plotted in Figure \ref{fig:figure14} in Appendix~\ref{AppA}.

Next, we consider the case of the intermediate tunneling range $\alpha=3$ in Fig. \ref{fig:figure4} (b1)-(b3) and the case of short range  tunneling $\alpha=5$ in panels (c1)-(c3).
Focusing on the latter case, the optimized dephasing profile follows an alternating pattern, with $\Gamma_n$ switching between small values ($\sim 10^{-5}$), to large values (up to 0.16), exceeding nearest neighbor energy gaps by about a factor of 1.5.
This pattern leads to a linearly-decaying population profile, see Fig. \ref{fig:figure4}(c2), in close agreement with the uniformly-optimized case.

One could rationalize that this pattern of dephasing broadens every second level to closely cover a three-site energy gap. In other words, instead of having every level broadened by $\Gamma_u$, every second level is broadened by $\approx 2 \Gamma_u$. 
The coherence map in Fig. \ref{fig:figure4}(c3)  shows that coherences are more extended when dephasing is optimized locally in this alternating manner. This contrasts with the population profile, which is similar whether or not optimization is done locally. 

The intermediate case $\alpha=3$ shows an in-between behavior between $\alpha=1$ and $\alpha=5$, with optimized $\Gamma_n$ values ranging around 0.05-0.10, 
see  Fig. \ref{fig:figure4}(b1).
The population, coherences, and flux of the site-optimized and uniform cases closely resemble each other.

In Appendix~\ref{AppB}, we repeat simulations with the energy gaps reduced to half of that used in Fig.~\ref{fig:figure4}, i.e., $\Delta = 0.5/N$, instead of $1/N$. As shown, general trends remain consistent with those reported in this section.

To quantify the spatial extent of coherences $\rho_{m\neq n}$, we define a coherence length (or coherence distance) as
\bea
\ell = 
\frac{\sum_{m \neq n} |m-n|\, |\rho_{mn}|}
{\sum_{m \neq n} |\rho_{mn}|}.
\label{eq:ell}
\eea
This measure corresponds to the average distance over which off-diagonal density matrix elements are distributed. Since this measure is basis dependent, we do not claim this coherence to be detectable or usable as a quantum resource, but it reflects extended delocalization. 

We define $\ell_{\mathrm{u}}$ as the coherence length under uniform dephasing, while $\ell_{\mathrm{o}}$ is the coherence length under the site-optimized dephasing.

\begin{figure}[htbp]
    \centering
\includegraphics[width=\linewidth]{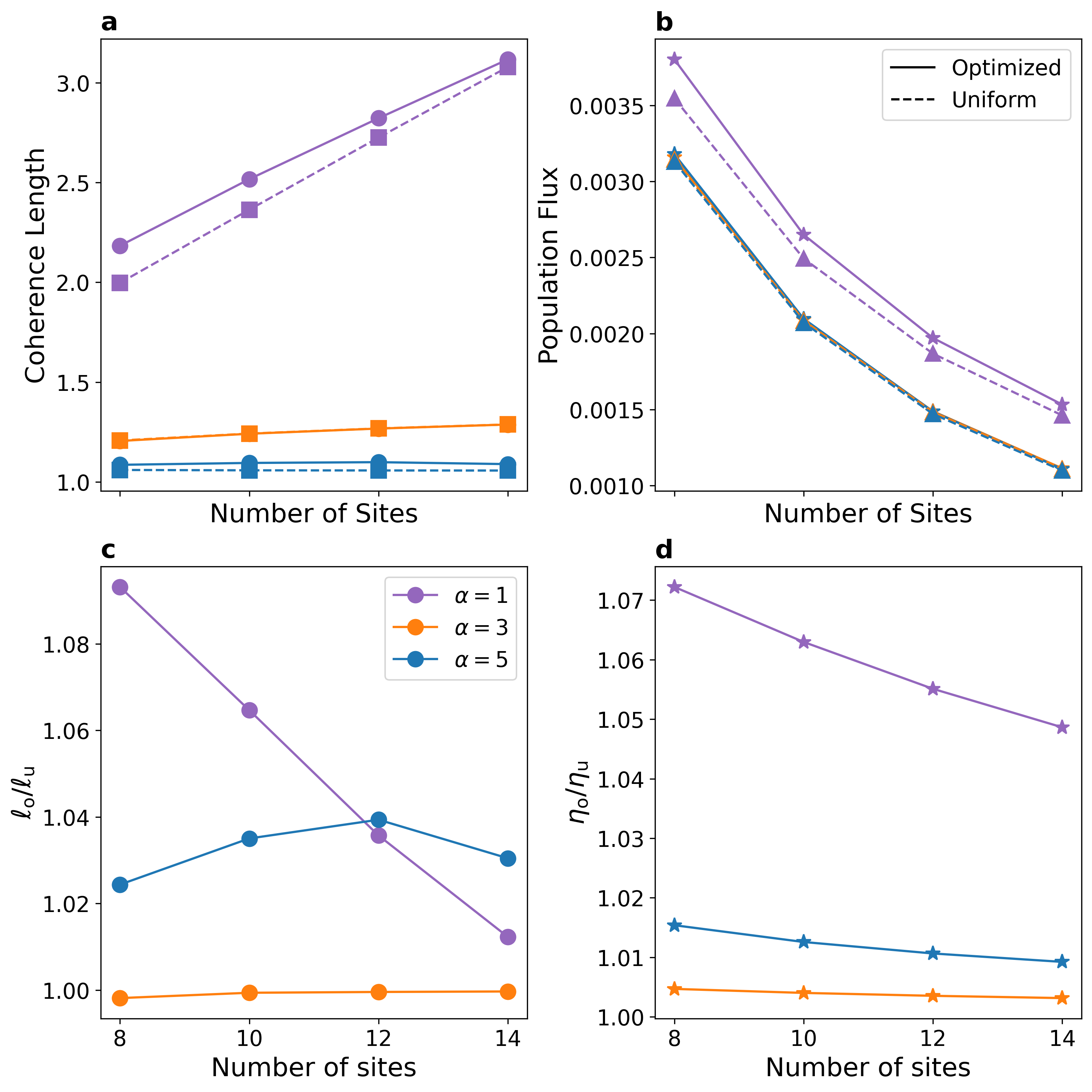}
\caption{
Analysis of coherence length and population flux in ramped systems with \(N = 8 - 14\) sites.
(a) Coherence length as a function of system length for optimized dephasing (solid lines, circles) and the best uniform dephasing (dashed lines, squares), for \(\alpha = 1\), \(\alpha = 3\), and \(\alpha = 5\) (purple, orange, and blue, respectively). 
(b) Population flux as a function of system length for optimized dephasing (solid lines, stars) and the best uniform dephasing (dashed lines, triangles). 
(c) Ratio of the coherence lengths under optimized and uniform dephasing. (d) Ratio of population fluxes under optimized and uniform dephasing rates. Constant energy differences between adjacent sites of $\Delta = 1/12$ were considered. Other parameters are \(J_{max} = 0.1\) and \(\gamma_l = 0.1\). }
\label{fig:figure5}
\end{figure}

We analyze this measure in Fig.~\ref{fig:figure5} along with the resulting flux for chains of 8, 10, 12, and 14 sites.  
For each length, the energy gap between adjacent sites was kept fixed at \(\Delta = 1/12\), and the best uniform dephasing rate, \(\Gamma_\mathrm{u}\), was calculated independently for each length. The optimal dephasing rates were found through the optimization process described in Sec. \ref{subsec:opt}. 
Overall, for each length, we find the site-optimized dephasing profile, the steady state, the coherence length, and the population flux, which we compare to the uniform dephasing scenario. 

First, in Fig.~\ref{fig:figure5}(a), we present the coherence length. For $\alpha=1$ the coherence length increases about linearly from around $\ell\approx2$ to $\ell \approx 3$ as the chain size grows from 8 to 14. This trend holds for both uniform and optimized cases. Similarly, the coherence length grows for $\alpha=3, 5$ with the size of system, but to a smaller extent. As for efficiency, we show in 
Fig.~\ref{fig:figure5}(b) that it decays with system size for both uniform and optimized cases. 

Moving on to the ratio of coherence lengths $\ell_{\mathrm{o}}/\ell_{\mathrm{u}}$ and the ratio of fluxes, $\eta_{\mathrm{o}}/\eta_{\mathrm{u}}$, both plotted as a function of system size, we  confirm that the efficiency ratio is consistently greater than unity.
Correspondingly, the coherence length is enhanced under site-optimized dephasing compared to the uniform case. 
Overall, the optimized setup yields more efficient transport compared to the uniform dephasing case by a few percent, accompanied by an increase in the delocalization of the steady state.

We summarize our observations regarding how locally-tuned dephasing can enhance transport under a ramp potential:

(i) {\bf Flux:} Engineering site-dependent dephasing allows a modest enhancement of transport flux (a measure for efficiency) relative to the uniformly-optimized case. This indicates that efficient population transfer in ramp-like quantum systems can be achieved without careful bath engineering. Nevertheless, under optimal site-dependent dephasing, the steady state supports longer range coherences than those observed in the uniform dephasing scenario.

(ii) {\bf Optimized local dephasing solutions:} WS localization can be disrupted by the introduction of dephasing, which effectively acts as level broadening. In the case of uniform dephasing, optimal transport is achieved when $\Gamma_{\rm u} \approx \Delta$, where $\Delta$ denotes the nearest-neighbor level spacing.
When optimizing dephasing locally, the tunneling power $\alpha$ dictates nontrivial behavior. For short-range tunneling, transport is optimized by increasing the dephasing strength to approximately $1.5\Delta$, but applied only to every second site. This setup promotes state delocalization.
In contrast, for long-range tunneling, the optimal strategy is to gradually increase the dephasing strength with distance from the injection site. This dephasing profile facilitates coherent tunneling to distant sites, followed by environment-assisted transport from those sites to the extraction site.

(iii) {\bf Nature of the steady state:} The site-optimized solution yields a steady state population profile, and thus population flux, that is very close in magnitude to the uniformly dephased case, with only a marginal improvement. 
However, the states are more delocalized under site-optimized dephasing. These enhanced coherences may be advantageous when considering figures of merit beyond flux, such as relaxation and dephasing times \cite{Tuokkola2025Transmon}.

(iv) {\bf Mechanism: } For short range tunneling, optimal solutions exhibit a population profile that decays approximately linearly with distance from the injection site, consistent with diffusive transport as described by Fick’s law. 
Steady state density matrix maps indicate that enhanced flux goes hand in hand with a higher level of quantum coherence. Applications that require extended state coherences \cite{Dutta2020,Plenio17,Kim2022PRA} may benefit from similar diffusive transport regimes. 


\begin{figure*}[htbp]
    \centering
\includegraphics[width=\linewidth]{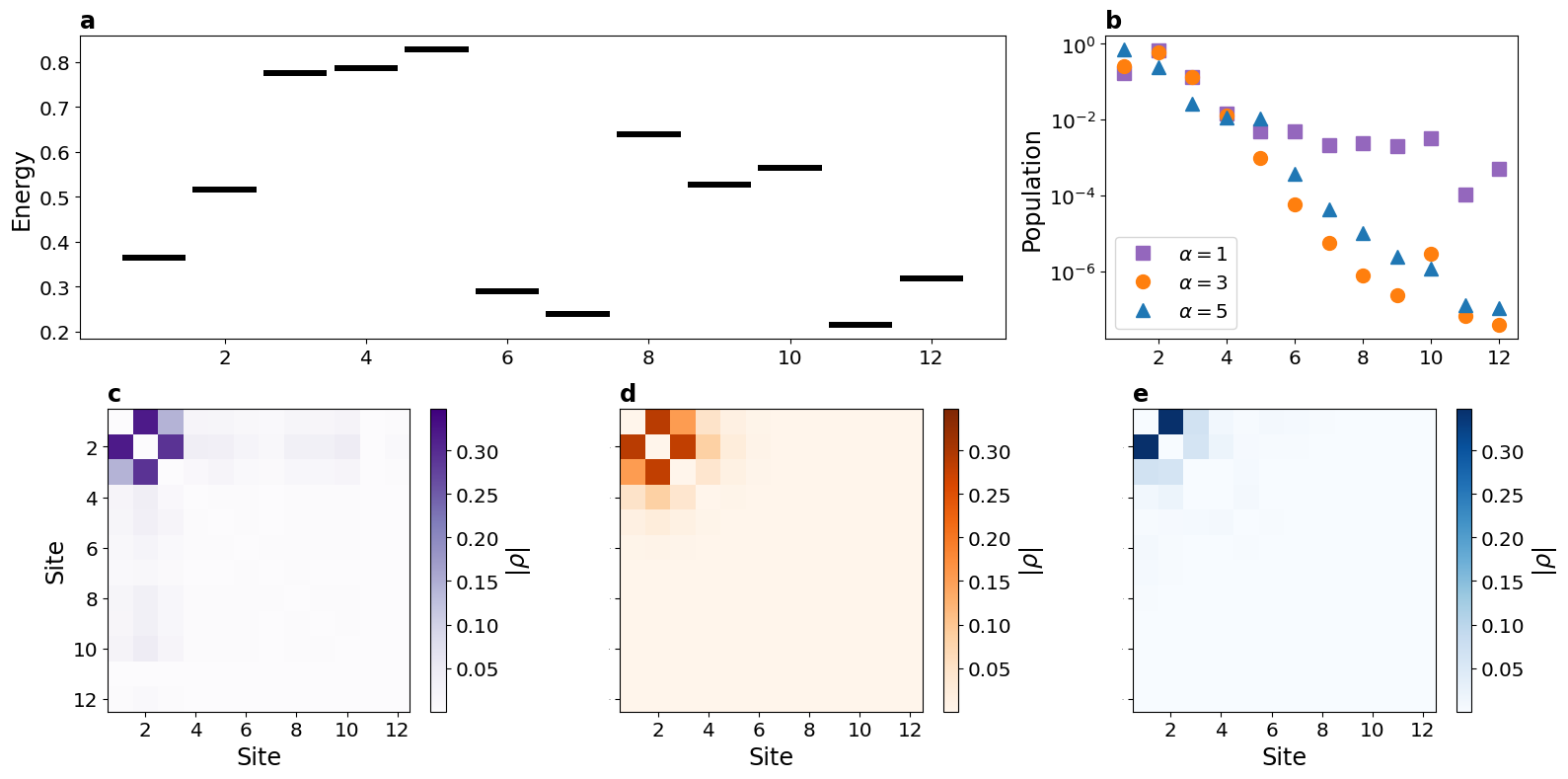}
    \caption{Development of Anderson localization in a disordered lattice of $N=12$ sites with no environmental interactions. (a) Example of a disordered energy profile. (b)   Steady-state population profiles for \(\alpha = 1, 3, 5\). All cases show suppression of population flux from the entrance site 1, but localization will strictly develop for $\alpha=5$. (c)-(e) steady-state density matrices for \(\alpha = 1,3,5\) with the diagonals removed. Other parameters are \(J_{max} = 0.1\) and \(\gamma_l = 0.1\). } 
\label{fig:figure6}
\end{figure*}

\begin{figure}[htbp]
    \centering
\includegraphics[width=0.8\linewidth] 
{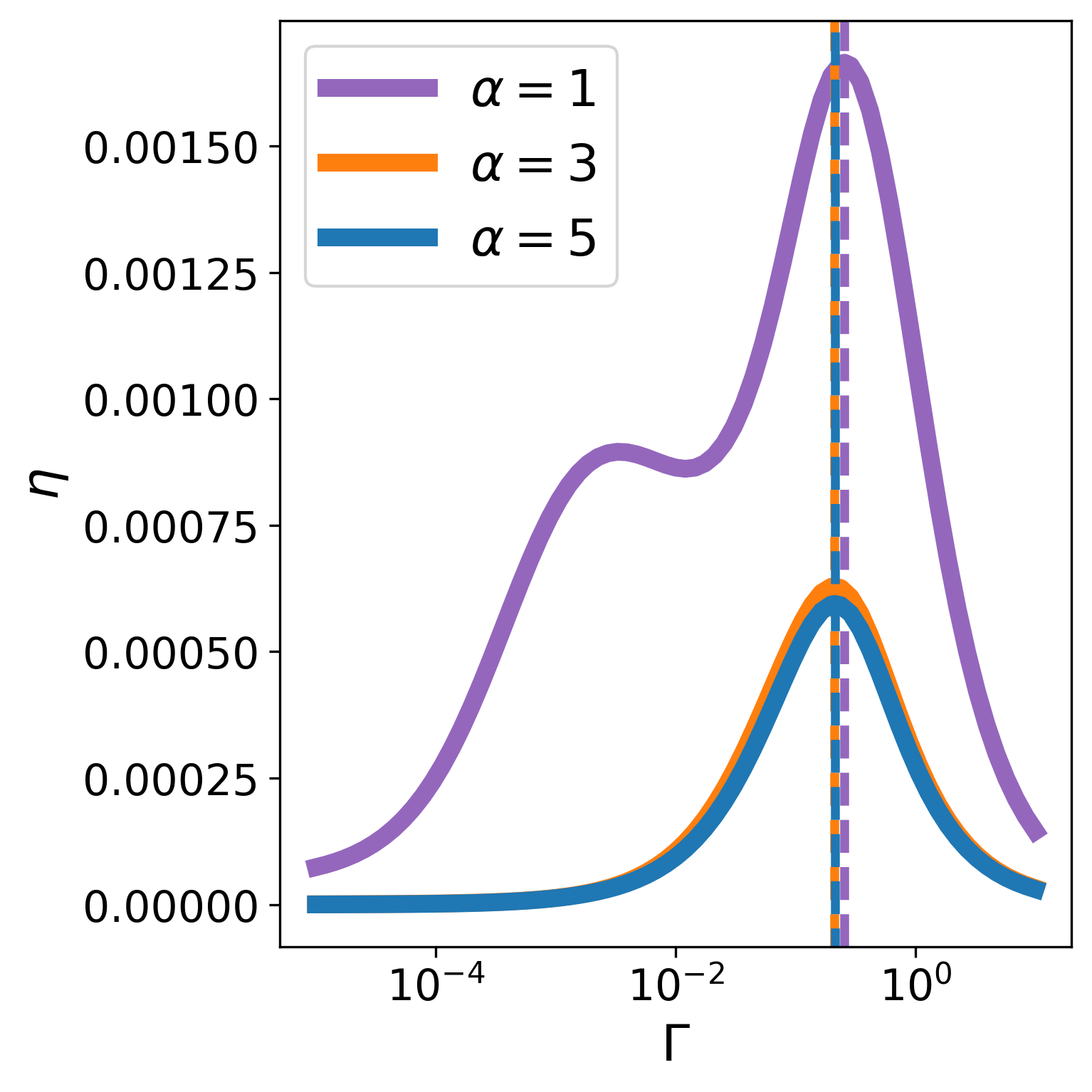}
    \caption{
Population flux as a function of uniform dephasing rate. 
The dashed lines show the locations of maximal flux: \(\Gamma_{\rm u} = \) 0.256, 0.211, and 0.214 with corresponding fluxes $\eta_{\alpha=1}=1.66 \cdot 10^{-3}$, $\eta_{\alpha=3}=6.29 \cdot 10^{-4}$, and $\eta_{\alpha=5}=5.94 \cdot 10^{-4}$, respectively.
Other parameters are the same as in Figure \ref{fig:figure6}.
} 
    \label{fig:figure7}
\end{figure}

\begin{figure*}
    \centering
\includegraphics[height=0.7\textheight]{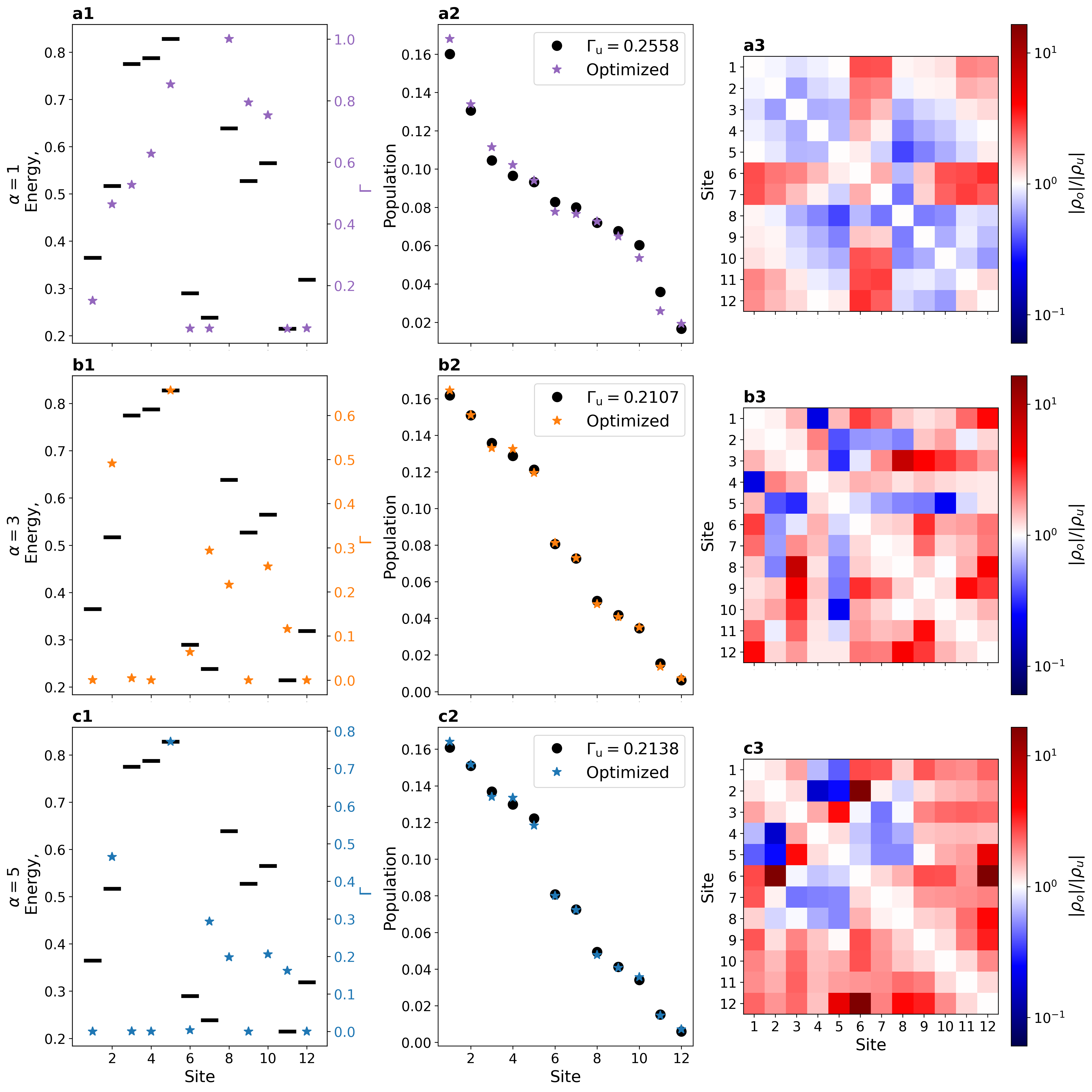}
    \caption{
Enhancing transport in an energetically disordered system with
site-optimized dephasing.    
\(\eta_{\alpha = 1} = 1.92 \cdot 10^{-3}\), \(\eta_{\alpha = 3} = 7.33 \cdot 10^{-4}\), \(\eta_{\alpha = 5} = 7.13 \cdot 10^{-4}\). Other parameters are the same as in Fig. \ref{fig:figure6}.
}  
    \label{fig:figure8}
\end{figure*}

\section{Overcoming Anderson Localization: Disordered Energy systems }
\label{sec:Anderson}

In one dimension, Anderson localization describes the striking result that for short-range tunneling potentials, even an arbitrarily small amount of disorder leads to exponential localization of wavefunctions and suppresses transport due to destructive interference.
\cite{anderson58}. As a consequence, nearest-neighbor one-dimensional models do not exhibit a diffusive or conducting phase, and transport is effectively absent at long distances.
Dephasing can enhance steady-state transport in disordered systems, as was shown in e.g., Refs. \cite{And10,And13}. 
These studies added a uniform dephasing to the system. Ref. \cite{AndR21} examined the case with randomly-placed onsite dephasing. Here, we continue with the exploration of site-optimized dephasing strength to further enhance transport.
We consider a finite system of 12 sites, observing signatures of localization physics. Within this setting, we investigate how locally-tuned dephasing can be used to optimize transport. 

\subsection{Representative example}

We begin with a disordered system in the absence of environmental effects. In Fig. \ref{fig:figure6}(a) we present an example of 12 sites with a strongly disordered energy profile, $\varepsilon \sim {\mathcal U}(0,1)$,
sampled independently for each site.

The resulting steady state populations are presented in  Fig. \ref{fig:figure6}(b) with coherences presented in panels (c)-(e) for $\alpha=1,3,5$, respectively.
When the tunneling energy is short-ranged with $\alpha=5$, the resulting population and state clearly show the development of localization physics. When the potential is longer-ranged ($\alpha=1,3$), we still see a strong decay of populations with distance, albeit slower, and the state remains delocalized. As our objective is generally to enhance transport rather than study localization physics, we do not explore in details the nature of the state in the coherent limit. 

We now introduce a uniform dephasing strength in the setup acting on each site, see Fig. \ref{fig:figure7}. Interestingly, we observe that while for $\alpha = 3,5$ the transport efficiency exhibits the standard single-peak ENAQT behavior, for the long-range case ($\alpha = 1$) the efficiency shows a double-peak structure as a function of $\Gamma$ \cite{Coates2023}. This reflects the presence of two distinct transport pathways, each optimized (and thus dominant) at a different dephasing range. We emphasize that this is a feature of this particular example and not a general trait observed for all disordered quantum networks.

We present the results of the optimization for local dephasing in Fig.~\ref{fig:figure8} considering the same disordered system as in Fig. \ref{fig:figure6}.
We study the long-range tunneling case in panels (a1)-(a3), the medium-range case in (b1)-(b3) and the short range scenario in (c1)-c3).
In each case, we find the set of $\Gamma_n$ that optimizes-enhances transport and we plot the population profile and coherences compared to simulations using the uniform value that optimized transport, $\Gamma_\mathrm{u}$.

We observe the following trends: 

(i) {\bf Optimized local dephasing solutions:} For both short-range and long-range tunneling, transport is optimized when dephasing is distributed non-uniformly across sites. The resulting profiles exhibit a non-periodic alternation between strong and weak local dephasing. 

(ii) {\bf Nature of the steady state:}
Inspection of the steady-state coherence map for long-range tunneling (panel a3) reveals that, under site-optimized dephasing, sites 1, 2, 6, 7, and 10–12 exhibit stronger mutual coherences than in the uniform dephasing case. Consistently, these sites experience relatively weak dephasing. Together, they form an effective sublattice that promotes transport: coherence is enhanced within this group, while remaining weaker between this sublattice and the rest of the chain, which forms a complementary set. More generally, under long-range tunneling, we observe that coherences preferentially develop among groups of states with similar energies, where dephasing is smallest.

A related trend is observed in the steady-state coherence map for short-range tunneling (panel c3). Under site-optimized dephasing, a spatially localized cluster (sites 6–12) emerges with enhanced coherence relative to the uniform dephasing case. This cluster is weakly coupled, in terms of coherence, to the first half of the chain. The two regions are effectively connected through the high-energy site 5, which experiences strong dephasing in the optimized configuration (see also Appendix~\ref{AppA}). These observations indicate that local optimization of dephasing can induce the formation of clusters of states with enhanced delocalization, thereby facilitating transport.

We thus find that optimal solutions tend to support more extended coherence than those obtained under uniform dephasing. In the long-range tunneling regime, transport is facilitated by clusters of sites that are quasi-resonant in energy. In contrast, for short-range tunneling, the relevant clusters are spatially localized.

\subsection{Characteristic trends: Ensemble averages}

We now consider an ensemble of 12-site systems with energy disorder, with the goal of identifying general principles governing how site-specific dephasing enhances transport, as well as the underlying mechanisms across different tunneling ranges.
Specifically, we generate 500 realizations, each characterized by independently sampled and uniformly distributed on-site energies satisfying $0 \leq \varepsilon_n \leq 1$. As an initial condition, all dephasing rates $\Gamma_n$ are set to a common value corresponding to the optimal transport observed in the uniform dephasing case, $\Gamma_{\mathrm{u}}$.
During the optimization, the dephasing rates are constrained within the range $10^{-7} \leq \Gamma_n \leq 1$.
In every instance, the optimized configuration yields transport performance that surpasses that of the uniform dephasing case.

\begin{figure}[htbp]
    \centering
    \includegraphics[width=\linewidth]{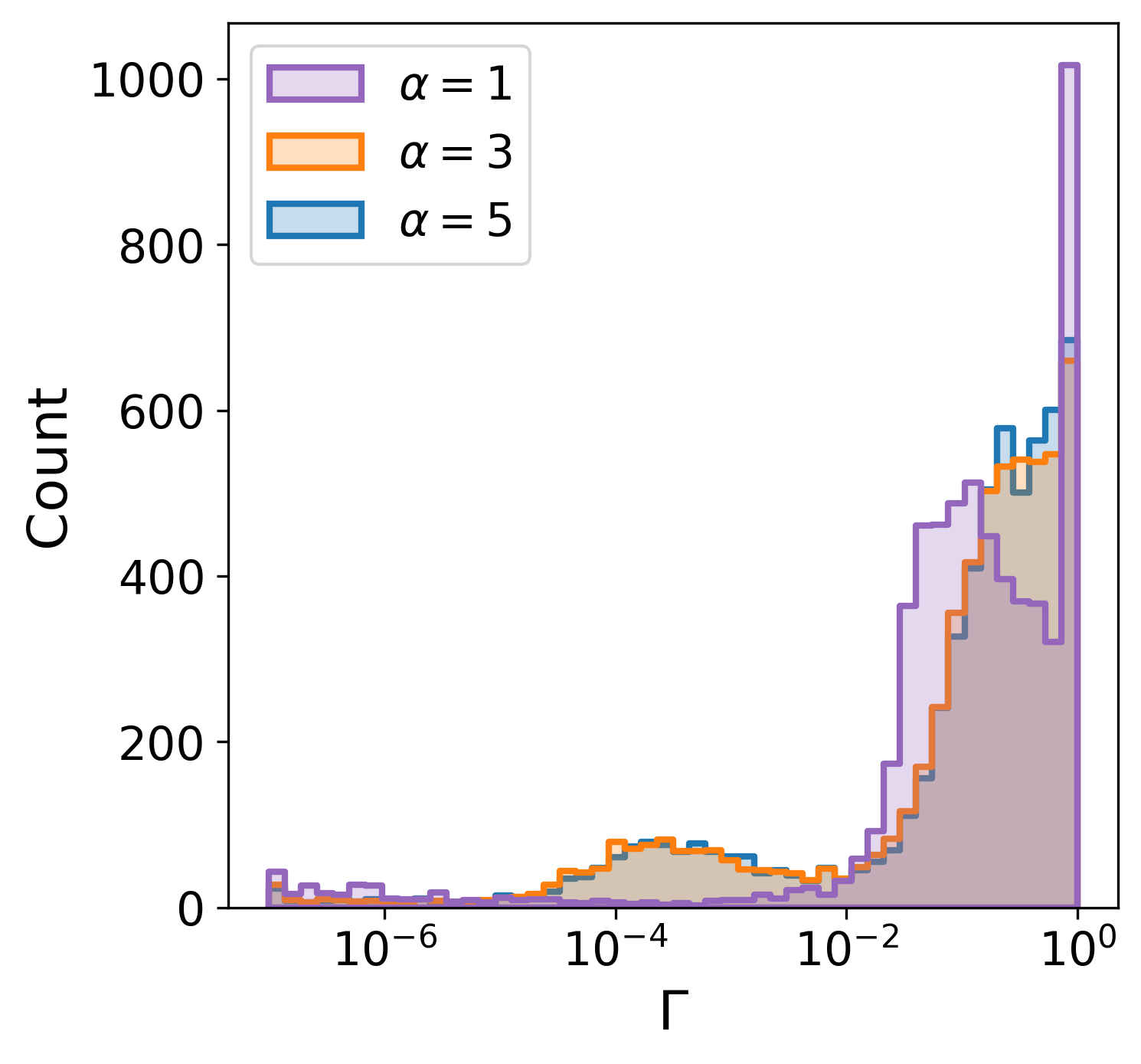}
    \caption{Distribution of all optimized \(\Gamma_n\) values in 500 \(N=12\) site disordered systems with \(\alpha = 1\) (purple), \(\alpha = 3\) (orange), and \(\alpha = 5\) (blue) power-law coupling. 
    } 
\label{fig:figure9}
\end{figure}


In Fig.~\ref{fig:figure9}, we present histograms of the optimized $\Gamma_n$ values across sites for the 500 cases. We first observe that, in many realizations, the optimized dephasing reaches the imposed upper bound. This suggests that even higher transport flux might be achievable if higher values of $\Gamma_n$ were allowed.
However, we set this upper bound as a physically motivated constraint: excessively large dephasing would lead to significant energy broadening, thereby obscuring the underlying energy landscape of the system and rendering site energies less meaningful.

Histograms exhibit distinct behavior for $\alpha = 1$ compared to $\alpha = 3,5$.
In the long-range regime, a significant fraction of the realizations attain the maximal allowed $\Gamma_n=1$, resulting in a strong level broadening, which is comparable to the overall range of $\varepsilon_n$. This behavior is consistent with transport dominated by long-range, near edge-to-edge transitions, which are optimized by substantial energy broadening.
In contrast, for short-range transport, two distinct clusters emerge: one at relatively strong dephasing, $0.1 \leq \Gamma_n \leq 1$, and another at weak dephasing, $\Gamma_n \approx 10^{-4}$. The high-dephasing cluster reflects the site-to-site nature of transport, where optimal performance is achieved when $\Gamma_n$ is on the order of the typical energy mismatch between neighboring sites. 
The low-dephasing cluster indicates that very small but nonzero dephasing improves transport by destroying destructive interference effects.

The individual case, Fig. \ref{fig:figure8}, 
suggested that site-optimized solutions sustain 
more extended coherences than when under uniform dephasing.  To quantify the spatial extent of coherences $\rho_{m\neq n}$, we employ the coherent length $\ell$ measure, Eq. (\ref{eq:ell}).

\begin{figure}[htbp]
    \centering
\includegraphics[width=\linewidth]{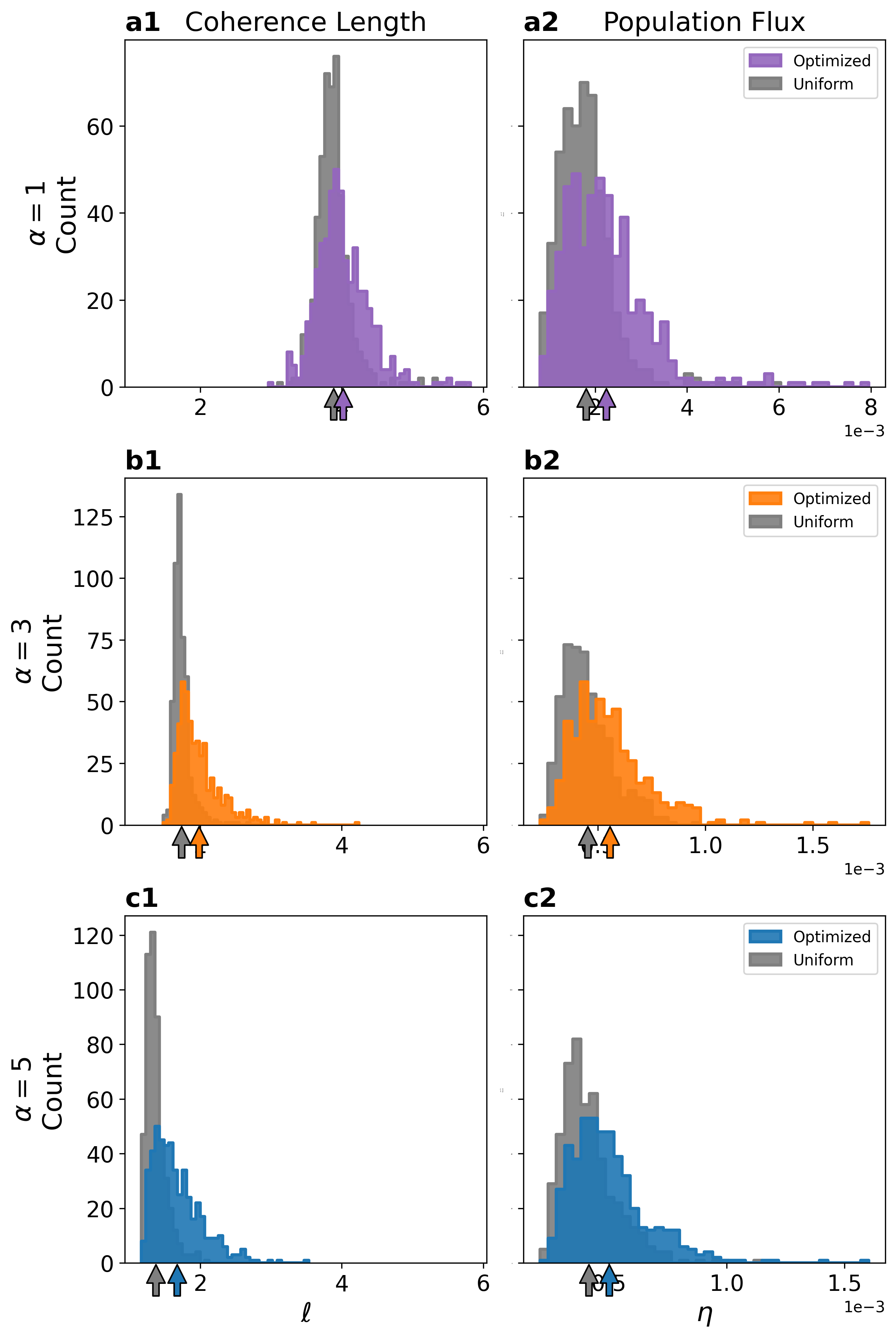}
    \caption{Histogram of coherence lengths (left) and population fluxes (right) for 500 disordered systems with \(\alpha = 1\) (a1-a2), \(\alpha = 3\) (b1-b2), and \(\alpha = 5\) (c1-c2) power-law tunneling. Results under uniform dephasing rates are shown in gray, and under the optimized dephasing rates in color. Arrows indicate mean values. The optimized-noise coherence lengths show average improvements over the uniform-noise coherence lengths of (a) 3.5\%, (b) 14.1\%, and (c) 21.5\%. The optimized-noise population fluxes show improvements over the uniform-noise population fluxes of (a) 21.9\%, (b) 21.8\%, and (c) 19.6\%.   }
\label{fig:figure10}
\end{figure}

In Fig.~\ref{fig:figure10}, we present histograms of the coherence length for uniform and site-optimized dephasing, $\ell_{\mathrm{u,o}}$, together with the corresponding population flux $\eta_{\mathrm{u,o}}$, for $\alpha = 1,3,5$. In all cases, the site-optimized configurations exhibit both larger coherence lengths and higher population fluxes compared to the uniform dephasing scenario.  
Furthermore, as expected, systems with long-range tunneling ($\alpha=1$) display the largest coherence lengths, reaching $\ell \approx 5$ sites. In contrast, short-range systems ($\alpha=5$) yield $\ell \approx 1.5$ even under site-optimized conditions.

\begin{figure*}[htbp]
    \centering
\includegraphics[width=0.8\linewidth]{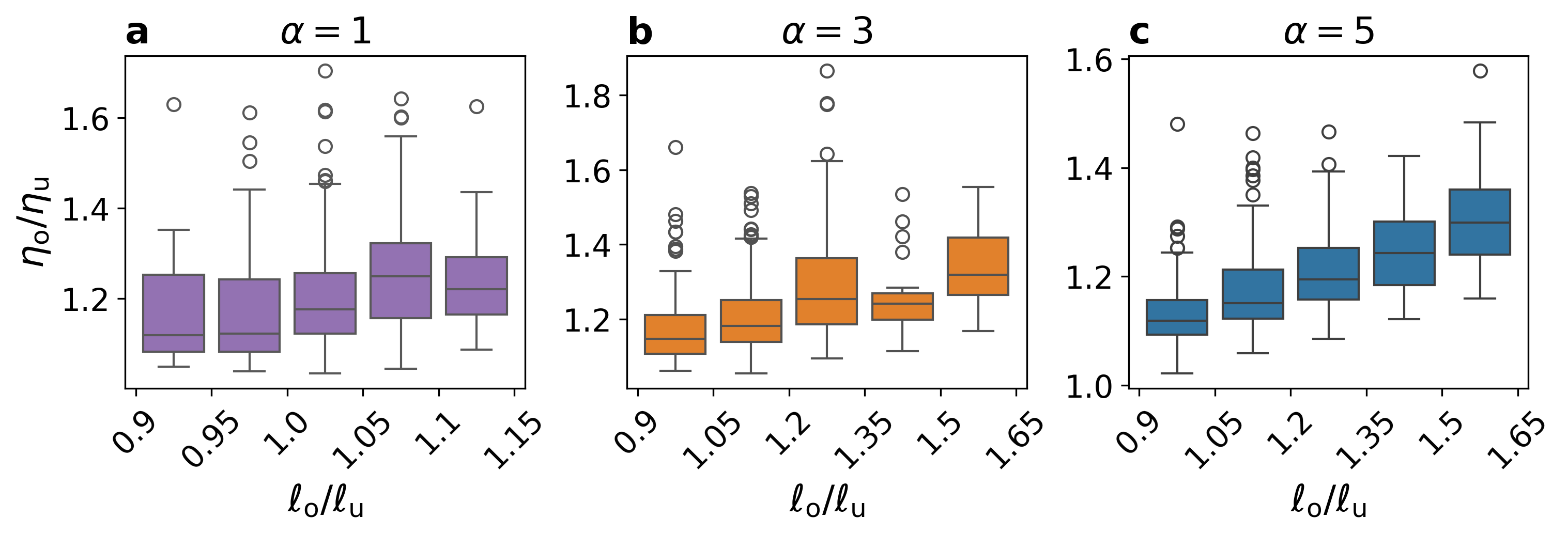}
    \caption{Boxplots of population flux improvements against binned coherence length ratios for disordered systems with power-law tunnelings of (a) \(\alpha = 1\), (b) \(\alpha = 3\), and (c) \(\alpha = 5\). The Spearman rank-order correlations are 
    (a) 0.33, (b) 0.45, and (c) 0.51.}
    \label{fig:figure11}
\end{figure*}

\begin{figure*}[htbp]
    \centering
\includegraphics[width=0.7\linewidth]{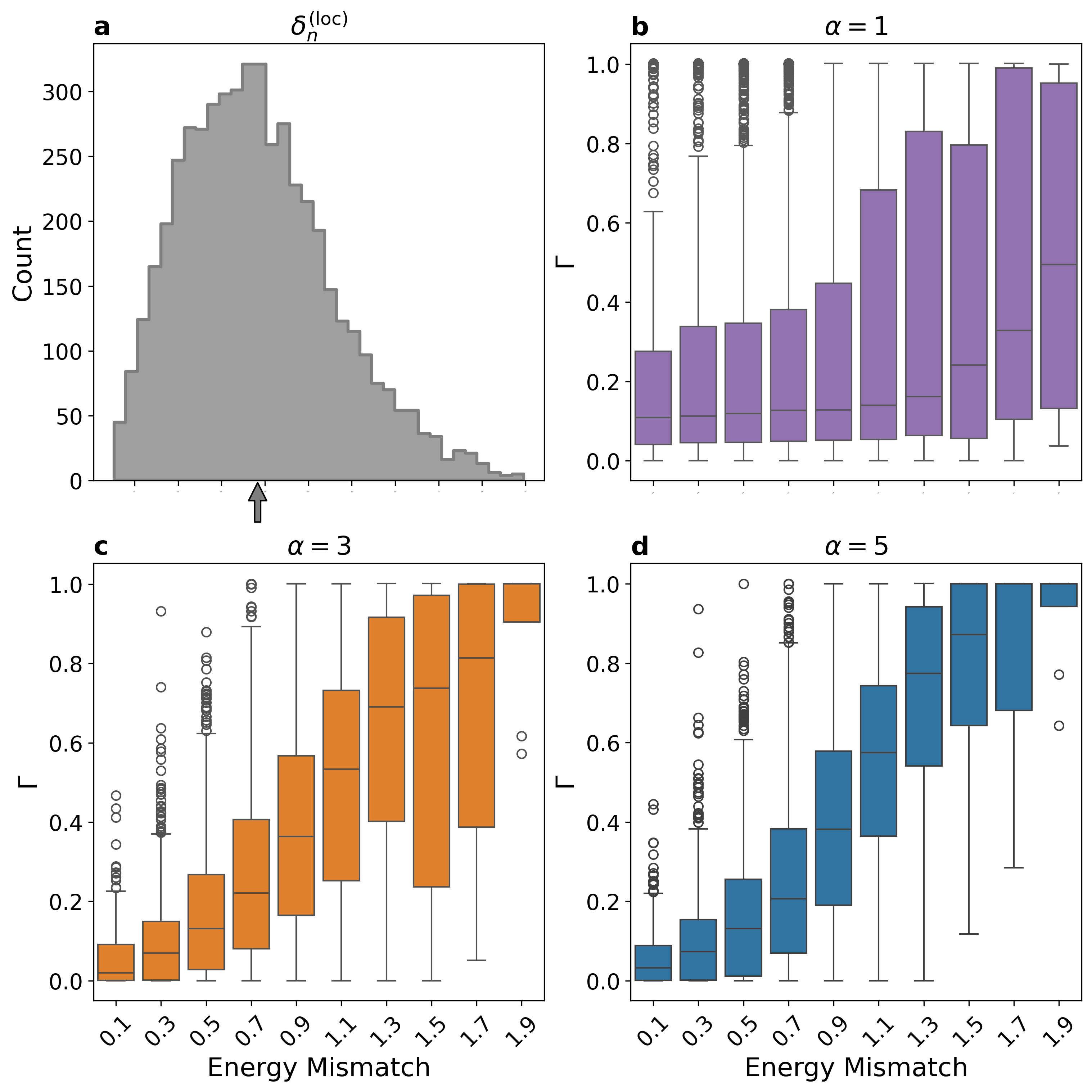}
    \caption{(a) Histogram of local energy mismatches $\delta_n^{(\mathrm{loc})}$, and boxplots of optimal \(\Gamma\) values as a function of binned local energy mismatches \(\delta_n^{(\mathrm{loc})}\) for (b) \(\alpha = 1\), (c) \(\alpha = 3\), and (d) \(\alpha = 5\) power-law tunneling. Plots generated over 500 samples. The x-label ticks indicate the centres of the boxplot bins. The Spearman rank-order correlations between dephasing rates and local energy mismatches are (a) 0.10, (b) 0.60, and (c) 0.66.}
    \label{fig:figure12}
\end{figure*}
We organize these results in Fig.~\ref{fig:figure11}, where, for each realization, we compare the average enhancement in flux with the corresponding change in coherence length. 
The colored boxes show the inter-quartile range (IQR), i.e. the range between the $25^{\mathrm{th}}$ and $75^{\mathrm{th}}$ percentiles of the data, with the median indicated by the horizontal line inside the box. The long whiskers extend to the last data point within $1.5 \times \mathrm{IQR}$ from the quartiles, and the circles mark values outside this range, considered to be statistical outliers \cite{MMSAboxplot}.

We find that for intermediate and short-range tunneling ($\alpha = 3,5$), the increase in relative flux correlates with an increase in the relative coherence length. This trend is less pronounced for $\alpha = 1$, where the enhancement in coherence due to site-optimized dephasing remains relatively small. We quantify these correlations using the Spearman rank-order correlation coefficient \cite{Spearman}. A Spearman \(r\) near 0 indicates there is no relationship between two random variables, positive values suggest a monotonic positive relationship, and negative values suggest a monotonic negative relationship. \(r\) is bounded by \(\pm1\). The results summarized in Fig. \ref{fig:figure11} have moderate Spearman rank-order correlations between population flux ratios and coherence length ratios of \(r = 0.33, 0.45, 0.51\) 
for \(\alpha = 1, 3, 5\), respectively, implying that improvements in flux will likely be accompanied by improvements in coherence length, and that this will occur most often for \(\alpha = 5\) short-range tunneling.
It is also worth noting that on average, in the short range tunneling scheme, dephasing optimization enhances transport by around 20\%. 


To gain insight into why particular site-dependent dephasing profiles are selected in disordered systems, we analyze correlations between the locally optimized dephasing, $\Gamma_n$, and local descriptors of the energy landscape, where we define the local energy mismatch as 
\bea
\delta_n^{(\mathrm{loc})}=|\varepsilon_n-\varepsilon_{n-1}|+|\varepsilon_n-\varepsilon_{n+1}|.
\eea
%
Across disordered system realizations, we calculate all local energy mismatches for the inner sites, bin them, and examine their corresponding distributions of optimized \(\Gamma_n\) using boxplots. We exclude the edge sites (1 and $N$) in this analysis to avoid having boundary effects mask underlying patterns in the bulk of the chain.

In Fig.~\ref{fig:figure12}, we present the resulting dependence of $\Gamma_n$ on the local energy mismatch $\delta_n^{(\mathrm{loc})}$. The colored boxes show the range between the $25^\mathrm{th}$ and $75^\mathrm{th}$ percentiles of the data, with the median indicated by the horizontal line inside the box. The long whiskers extend to the last data point within $1.5 \times \mathrm{IQR}$ from the quartiles; circles mark values outside this range. 
For $\alpha = 1$, we observe no correlations between $\Gamma_n$ and $\delta_n^{(\mathrm{loc})}$, in line with the understanding that by definition, this system supports long range tunneling. In contrast, for $\alpha = 3,5$, more significant relationships emerge, with Spearman rank-order correlations of \(r = 0.560, 0.66\), respectively. This means that $\Gamma_n$ increases are often accompanied by increases of $\delta_n^{(\mathrm{loc})}$.
This behavior can be understood from the local nature of transport in the short-range regime, where larger dephasing is required to overcome significant local energy mismatches.
Note that very few samples are used to draw the right-most boxplots, as shown by panel (a), leading in some cases to the median nearly overlapping with one of the other quartiles.

Overall, we find that in finite-size energy disordered systems, transport is optimized when dephasing is applied non-uniformly across sites. In particular, allowing only a subset of sites to experience dephasing leads to states that are more delocalized compared to the uniform dephasing scenario. We also find that, in agreement with our intuition, larger energy mismatches between neighboring sites correlate with stronger local dephasing in the short-range tunneling regime.


\begin{figure}[htbp]
    \centering
\includegraphics[width=\linewidth]{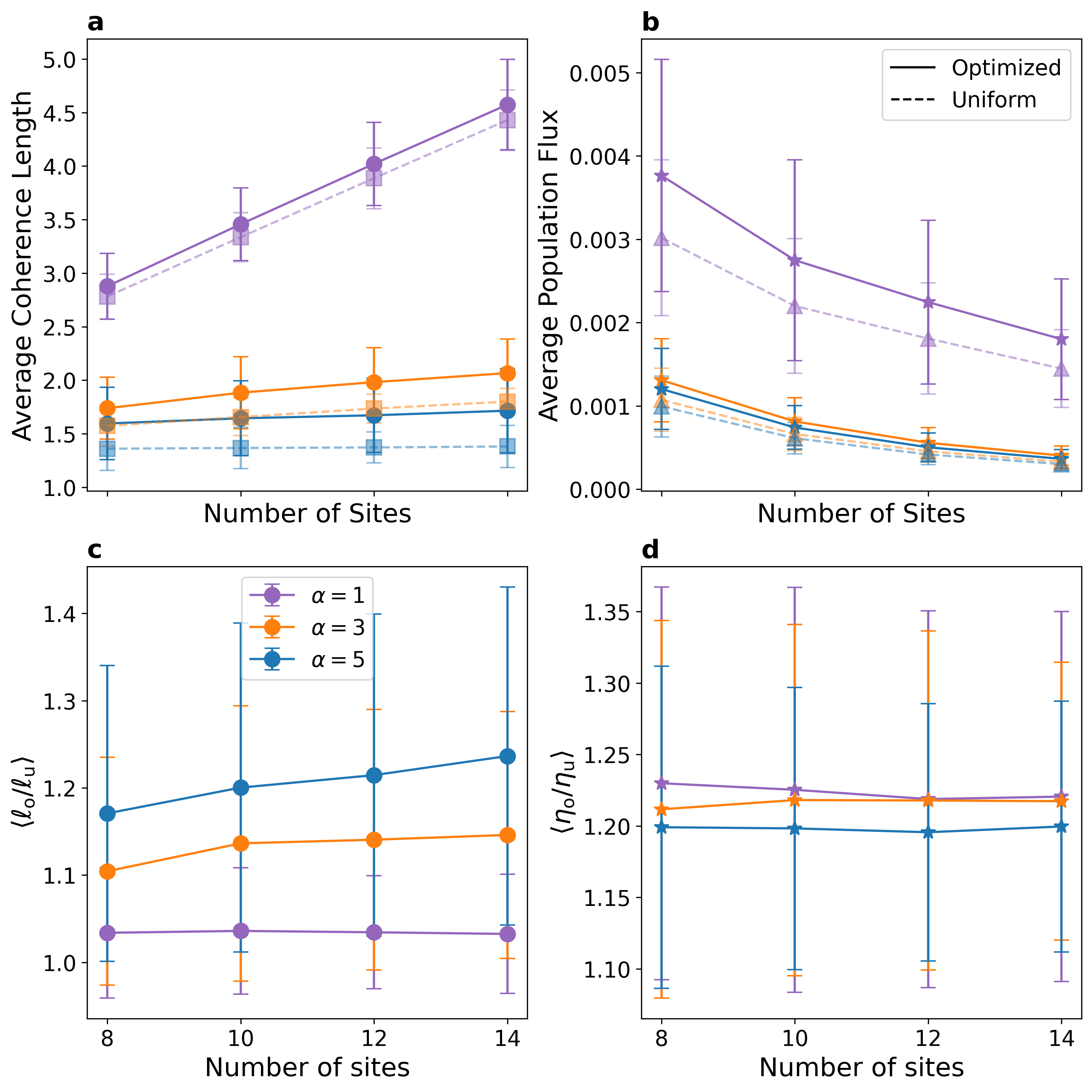}
\caption{
Analysis of coherence length and flux in disordered systems with $N = 8-14$ sites.
(a) Average coherence length as a function of system size for site-dependent optimized dephasing rates (solid lines with circles) and for the optimal uniform dephasing rate (dashed lines with squares). Results are shown for $\alpha = 1, 3, 5$ (purple, orange, and blue, respectively).
(b) Average population flux as a function of system length for optimized dephasing rates (solid lines with stars) and the best uniform dephasing rate (dashed lines, triangles). 
(c) Average of the ratios of the coherence lengths under optimized and uniform dephasing rates. 
(d) Average of the ratios of population fluxes under optimized and uniform dephasing rates. 
Averages were calculated over 500 independently-generated disordered systems for each system length. Disordered site energies were generated between 0 and 1. Other parameters are \(J_{max} = 0.1\) and \(\gamma_l = 0.1\). The error bars are calculated as the standard deviations of plotted quantities.}
\label{fig:figure13}
\end{figure}

To further substantiate that site-optimized dephasing enhances coherence relative to the uniform case, we study chains of varying length. Figure~\ref{fig:figure13} analyzes the average coherence length and population flux in disordered systems with \(N = 8, 10, 12, 14\) sites, where 500 realizations are generated for each system size. 
Panel \ref{fig:figure13}(a) shows the average coherence length for optimized and uniform dephasing rates. For all three tunneling regimes, the coherence length increases with system size, with the most pronounced growth observed for long-range tunneling. As expected and consistent with the ramp-system results in Fig.~\ref{fig:figure5}, the population flux decreases monotonically as additional sites are introduced, as shown in Panel~\ref{fig:figure13}(b). 
Panels \ref{fig:figure13}(c) and (d) show the relative improvements in coherence length and population flux, respectively, obtained using site-optimized dephasing compared to the uniform case. For long-range tunneling, the enhancement in coherence length remains nearly constant as the system size increases. In contrast, for shorter-range tunneling (\(\alpha = 3\) and \(\alpha = 5\)), the enhancement ratio in coherence length exhibits a gradual increase with system size. 
Overall, optimizing the dephasing rates  yields an approximately \(20\%\) increase in population flux across all system sizes considered, compared to the case with a uniform optimal dephasing rate \(\Gamma_u\).


\section{Summary}
\label{sec:Summ}

In this work, we investigated whether locally engineered environmental noise can enhance steady-state single-particle one-dimensional quantum transport, and what mechanisms support it. We focused on finite-size systems and steady-state transport. Our approach relied on a local Lindblad quantum master equation to simulate quantum dynamics under environmental effects, and we optimized site-dependent dephasing constants to maximize the population flux. 

We considered two types of finite-size systems that experience localization in the thermodynamic limit for short-range tunneling: systems under a ramp potential (Wannier-Stark localization) and  systems with static energy disorder (Anderson localization). Our analysis combined analytical results for minimal three-site models with numerical optimization for extended systems. It allowed us to identify design principles for site-engineered environmental noise-assisted quantum transport.

\begin{table*}[!htbp]
\centering
\caption{Design principles for optimizing site-dependent dephasing in quantum transport.}
\label{tab:tableS}
\begin{tabular}{p{2cm} p{2.6cm} p{4.5cm} @{\hspace{0.7cm}} p{7.5cm}}
\hline\hline
\textbf{Model} & \textbf{Tunneling regime} & \textbf{Optimal dephasing profile $\Gamma_n$} & \textbf{Mechanism and key insight} \\
\hline

Ramp 
& Short-range  
& Alternating high and low dephasing 
& Selective level broadening enhances spectral overlap between neighboring sites. 
Only a subset of sites requires dephasing to restore transport. \\

Ramp 
& Long-range
& Dephasing increases with distance from injection site 
& Dephasing compensates growing energy detuning across the chain.  
The growing pattern maintains coherences near the source and creates spectral overlap to distant sites. \\

Disordered 
& Short-range 
& Strongly non-uniform profile 
& Local dephasing overcomes local energy mismatch;
$\Gamma_n$ correlates with local energy mismatch $\delta_n^{(\mathrm{loc})}$ \\

Disordered 
& Long-range 
& Broad distribution; some sites require large dephasing 
& Strong level broadening enables long-range hopping between distant sites; weak correlations observed between local dephasing and local energy mismatch. \\

\hline\hline
\end{tabular}
\end{table*}

\subsection{Design Principles}

Spatially structured environmental noise generally provides a physically transparent route to controlling quantum transport. Rather than uniformly suppressing coherence, optimal strategies balance coherence preservation and dephasing-induced level broadening across the system. While the resulting improvements in population flux are modest in ordered systems, they become more significant in disordered settings and are consistently accompanied by enhanced state delocalization. 

As a general guideline, we find that dephasing acts as an effective spectral broadening that matches energy mismatches across transport pathways. Optimal transport emerges from balancing where to destroy coherence to overcome energy mismatch, and where to preserve it to enable delocalization.
In short-range systems, this broadening enabled intermediate sites to bridge energy gaps between neighbors, while in long-range systems, it facilitated direct transitions between distant sites by bringing them into resonance. 

For systems with a ramp-like energy landscape, we demonstrated that the optimal dephasing profile depended on the range of tunneling. In the short-range regime, transport was enhanced by selectively applying dephasing to a subset of sites, effectively broadening alternating energy levels. In contrast, for long-range tunneling, optimal transport was achieved by gradually increasing the dephasing strength along the chain, promoting long-distance transitions while mitigating localization induced by the energy gradient. In all cases, the locally-optimized dephasing profiles resulted in states that were more delocalized than states generated under the optimal uniform dephasing.

In energy disordered systems, we found that the optimized dephasing profiles were generally heterogeneous, leading to more delocalized states than those obtained under optimal-uniform dephasing. In the short-range tunneling regime, strongly-detuned sites tended to require enhanced local dephasing to facilitate transport. In contrast, such correlations were absent in long-range tunneling models.


Within our parameter scheme, dephasing optimization enhanced the flux by a few percent only for the ramp potential. In contrast, for systems with energy disorder, there was around a 20\% enhancement in flux by optimizing dephasing locally. 
Furthermore, site-optimized dephasing consistently produced states with more extended spatial coherence (compared to the uniform case), or states with clusters of enhanced coherence in the disordered regime. Environmental engineering can thus be used to control the extent of quantum coherence in the system, relevant for applications where coherence play a functional role beyond transport.
Studying these effects in longer systems remains an open question, as is understanding whether the transition to diffusive transport is distinct under uniform or site-optimized dephasing. 

We organize these observations, listing design principles for locally-engineered environmental noise in Table \ref{tab:tableS}.

\subsection{Future work}

In molecular systems, spatial variations in the environment arise naturally, for example, as a result of differences in local vibrational couplings. Such heterogeneity can give rise to effective non-uniform dephasing across sites \cite{Plenio08}. 
Our work can be tested in engineered realizations of site-dependent dephasing. This can be done, e.g., with trapped-ion systems, where local dephasing is achieved by applying site-specific noise through laser-induced fluctuations \cite{maier2019environment}. Similarly, superconducting qubit arrays allow for local tunable dissipation and dephasing via engineered environments and local control lines \cite{sundelin2026quantumrefrigeration}. 
Photonic lattices provide another route, where effective dephasing can be introduced through controlled disorder or coupling to auxiliary lossy modes \cite{photon24, biggerstaff2016waveguides}.


This study was focused on systems inspired by occurrences of Wannier-Stark and Anderson localization in the thermodynamic limit, and on the impact of both uniform and locally-engineered dephasing in breaking these localization mechanisms.  More generally, dephasing is known to enhance transport in other classes of otherwise localized systems. Examples include the time-dependent Anderson Hamiltonian~\cite{TimeD90}, as well as quasiperiodic systems such as the Aubry--André--Harper and Fibonacci models~\cite{Landi21}. Investigating the role of site-engineered dephasing in such systems is an interesting direction for future work.
Enhancing quantum dissipative transport under many-body interactions remains a challenging computational problem \cite{MBLD16a,MBLD16b,MBLD16c,bijay25}.
As this study focused on single-particle transport, future efforts will explore enhancing transport through dephasing engineering in multi-excitonic many-body systems.

\vspace{4mm}
\begin{acknowledgments}
The work of M.L. is  supported by the NSERC Canada Graduate Scholarship-Doctoral. E. W. was supported by the NSERC Undergraduate Research Award (USRA). 
D.S. acknowledges support from an NSERC Discovery Grant and an NSERC Alliance International Catalyst Grant. 
Resources used in preparing this research were provided, in part, by the Province of Ontario, the Government of Canada through CIFAR, and companies sponsoring the Vector Institute \url{www.vectorinstitute.ai/#partners}. 
\end{acknowledgments}

\bibliography{references}

@article{anderson58,
  title = {Absence of diffusion in certain random lattices},
  author = {Anderson, P.~W.},
  journal = {Phys. Rev.},
  volume = {109},
  pages = {1492--1505},
  year = {1958},
  doi = {10.1103/PhysRev.109.1492}
}

@article{emin87,
  title = {Existence of Wannier-Stark localization},
  author = {Emin, David and Hart, C. F.},
  journal = {Phys. Rev. B},
  volume = {36},
  issue = {14},
  pages = {7353--7359},
  numpages = {0},
  year = {1987},
  month = {Nov},
  publisher = {American Physical Society},
  doi = {10.1103/PhysRevB.36.7353},
  url = {https://link.aps.org/doi/10.1103/PhysRevB.36.7353
        }
}

@article{PL99,
  title = {Experimental Observation of Linear and Nonlinear Optical Bloch Oscillations},
  author = {Morandotti, R. and Peschel, U. and Aitchison, J. S. and Eisenberg, H. S. and Silberberg, Y.},
  journal = {Phys. Rev. Lett.},
  volume = {83},
  issue = {23},
  pages = {4756--4759},
  numpages = {0},
  year = {1999},
  month = {Dec},
  publisher = {American Physical Society},
  doi = {10.1103/PhysRevLett.83.4756},
  url = {https://link.aps.org/doi/10.1103/PhysRevLett.83.4756}
}

@article{OL07,
  author  = {Tal Schwartz and Guy Bartal and Shmuel Fishman and Mordechai Segev},
  title   = {Transport and Anderson localization in disordered two-dimensional photonic lattices},
  journal = {Nature},
  year    = {2007},
  volume  = {446},
  pages   = {52--55},
  doi     = {10.1038/nature05623}
}

@article{SC22,
  title = {Quantum transport and localization in 1d and 2d tight-binding lattices},
  author = {Karamlou, Amir H. and Braum{\"u}ller, Jochen and Yanay, Yariv and Di Paolo, Agustin and Harrington, Patrick M. and Kannan, Bharath and Kim, David and Kjaergaard, Morten and Melville, Alexander and Muschinske, Sarah and Niedzielski, Bethany M. and Veps{\"a}l{\"a}inen, Antti and Winik, Roni and Yoder, Jonilyn L. and Schwartz, Mollie and Tahan, Charles and Orlando, Terry P. and Gustavsson, Simon and Oliver, William D.},
  journal = {npj Quantum Information},
  volume = {8},
  pages = {35},
  year = {2022},
  doi = {10.1038/s41534-022-00528-0}
}

@misc{deepmind2020jax,
  title = {The {D}eep{M}ind {JAX} {E}cosystem},
  author = {DeepMind and Babuschkin, Igor and Baumli, Kate and Bell, Alison and Bhupatiraju, Surya and Bruce, Jake and Buchlovsky, Peter and Budden, David and Cai, Trevor and Clark, Aidan and Danihelka, Ivo and Dedieu, Antoine and Fantacci, Claudio and Godwin, Jonathan and Jones, Chris and Hemsley, Ross and Hennigan, Tom and Hessel, Matteo and Hou, Shaobo and Kapturowski, Steven and Keck, Thomas and Kemaev, Iurii and King, Michael and Kunesch, Markus and Martens, Lena and Merzic, Hamza and Mikulik, Vladimir and Norman, Tamara and Papamakarios, George and Quan, John and Ring, Roman and Ruiz, Francisco and Sanchez, Alvaro and Sartran, Laurent and Schneider, Rosalia and Sezener, Eren and Spencer, Stephen and Srinivasan, Srivatsan and Stanojevi\'{c}, Milo\v{s} and Stokowiec, Wojciech and Wang, Luyu and Zhou, Guangyao and Viola, Fabio},
  url = {http://github.com/google-deepmind},
  year = {2020},
}

@article{Blach25,
  author       = {Blach, D. D. and Lumsargis-Roth, V. A. and Chuang, C. and others},
  title        = {Environment-assisted quantum transport of excitons in perovskite nanocrystal superlattices},
  journal      = {Nature Communications},
  volume       = {16},
  pages        = {1270},
  year         = {2025},
  doi          = {10.1038/s41467-024-55812-8},
  url          = {https://doi.org/10.1038/s41467-024-55812-8}
}

@article{maier2019environment,
  title = {Environment-Assisted Quantum Transport in a 10-qubit Network},
  author = {Christine Maier and Tiff Brydges and Petar Jurcevic and Nils Trautmann and Cornelius Hempel and Ben P. Lanyon and Philipp Hauke and Rainer Blatt and Christian F. Roos},
  journal = {Physical Review Letters},
  volume = {122},
  number = {5},
  pages = {050501},
  year = {2019},
  doi = {10.1103/PhysRevLett.122.050501}
}

@incollection{breuer2007_decoherence,
  author       = {Heinz-Peter Breuer and Francesco Petruccione},
  title        = {Decoherence},
  booktitle    = {The Theory of Open Quantum Systems},
  year         = {2007},
  publisher    = {Oxford University Press},
  address      = {Oxford},
  chapter      = {Decoherence},
  isbn         = {9780199213900},
  doi          = {10.1093/acprof:oso/9780199213900.003.04},
  pages        = {219-282}
}

@article{juhasz2018vibrations,
  author       = {Imre Benedek Juhász and Árpád I. Csurgay},
  title        = {Impact of undamped and damped intramolecular vibrations on the efficiency of photosynthetic exciton energy transfer},
  journal      = {AIP Advances},
  volume       = {8},
  number       = {4},
  pages        = {045318},
  year         = {2018},
  month        = {April},
  doi          = {10.1063/1.5009114},
  url          = {https://doi.org/10.1063/1.5009114},
  note         = {Open Access}
}

@article{Jager2022,
  title = {Lindblad Master Equations for Quantum Systems Coupled to Dissipative Bosonic Modes},
  author = {J\"ager, Simon B. and Schmit, Tom and Morigi, Giovanna and Holland, Murray J. and Betzholz, Ralf},
  journal = {Phys. Rev. Lett.},
  volume = {129},
  issue = {6},
  pages = {063601},
  numpages = {7},
  year = {2022},
  month = {Aug},
  publisher = {American Physical Society},
  doi = {10.1103/PhysRevLett.129.063601},
  url = {https://link.aps.org/doi/10.1103/PhysRevLett.129.063601}
}

@article{giri2025entanglement,
  author       = {Sajal Kumar Giri and George C. Schatz},
  title        = {Modeling entanglement dynamics of molecules interacting with entangled photons through Lindblad master equation approach},
  journal      = {The Journal of Chemical Physics},
  volume       = {162},
  number       = {11},
  pages        = {114106},
  year         = {2025},
  month        = {March},
  doi          = {10.1063/5.0254272},
  url          = {https://doi.org/10.1063/5.0254272},
  note         = {Special Collection: Festschrift for Abraham Nitzan}
}

@INPROCEEDINGS{Rosati2014,
  author={Rosati, Roberto and Iotti, Rita Claudia and Rossi, Fausto},
  booktitle={2014 International Workshop on Computational Electronics (IWCE)}, 
  title={Microscopic modeling of quantum devices at high carrier densities via Lindblad-type scattering superoperators}, 
  year={2014},
  volume={},
  number={},
  pages={1-3},
  doi={10.1109/IWCE.2014.6865848}
}

@article{sarkar2020environment,
  author    = {Subhajit Sarkar and Yonatan Dubi},
  title     = {Environment-Assisted and Environment-Hampered Efficiency at Maximum Power in a Molecular Photocell},
  journal   = {The Journal of Physical Chemistry C},
  volume    = {124},
  number    = {28},
  pages     = {15115--15122},
  year      = {2020},
  publisher = {American Chemical Society},
  doi       = {10.1021/acs.jpcc.0c04581},
  url       = {https://dx.doi.org/10.1021/acs.jpcc.0c04581?ref=pdf}
}

@article{Dutta2021OutOfEquilibrium,
  title = {Out-of-equilibrium steady states of a locally driven lossy qubit array},
  author = {Dutta, Shovan and Cooper, Nigel R.},
  journal = {Phys. Rev. Res.},
  volume = {3},
  issue = {1},
  pages = {L012016},
  numpages = {7},
  year = {2021},
  month = {Feb},
  publisher = {American Physical Society},
  doi = {10.1103/PhysRevResearch.3.L012016},
  url = {https://link.aps.org/doi/10.1103/PhysRevResearch.3.L012016}
}

@article{manzano2013quantum,
  author  = {Manzano, Daniel},
  title   = {Quantum Transport in Networks and Photosynthetic Complexes at the Steady State},
  journal = {PLOS ONE},
  year    = {2013},
  volume  = {8},
  number  = {2},
  pages   = {e57041},
  doi     = {10.1371/journal.pone.0057041}
}

@article{ML25,
  title = {Discovery of energy landscapes toward optimized quantum transport: Environmental effects and long-range tunneling},
  author = {Lawrence, Maggie and Pocrnic, Matthew and Fung, Erin and Carrasquilla, Juan and Gauger, Erik M. and Segal, Dvira},
  journal = {Phys. Rev. Res.},
  volume = {8},
  issue = {1},
  pages = {013163},
  numpages = {26},
  year = {2026},
  month = {Feb},
  publisher = {American Physical Society},
  doi = {10.1103/ckfp-6jtd},
  url = {https://link.aps.org/doi/10.1103/ckfp-6jtd}
}

@article{Alan08,
  title = {Environment-assisted quantum walks in photosynthetic energy transfer},
  author = {Mohseni, Masoud and Rebentrost, Patrick and Lloyd, Seth and Aspuru-Guzik, Alán},
  journal = {The Journal of Chemical Physics},
  volume = {129},
  number = {17},
  pages = {174106},
  year = {2008},
  doi = {10.1063/1.3002335},
  url = {https://doi.org/10.1063/1.3002335}
}

@article{Plenio21,
  author       = {Andrea Mattioni and Felipe Caycedo-Soler and Susana F. Huelga and Martin B. Plenio},
  title        = {Design Principles for Long-Range Energy Transfer at Room Temperature},
  journal      = {Physical Review X},
  volume       = {11},
  pages        = {041003},
  year         = {2021},
  doi          = {10.1103/PhysRevX.11.041003}
}

@article{Plenio12,
  author       = {A. W. Chin and Susana F. Huelga and Martin B. Plenio},
  title        = {Coherence and decoherence in biological systems: principles of noise-assisted transport and the origin of long-lived coherences},
  journal      = {Philosophical Transactions of the Royal Society A: Mathematical, Physical and Engineering Sciences},
  volume       = {370},
  number       = {1972},
  pages        = {3638--3657},
  year         = {2012},
  publisher    = {The Royal Society Publishing},
  doi          = {10.1098/rsta.2011.0224}
}

@article{Plenio10,
  author       = {A. W. Chin and A. Datta and F. Caruso and S. F. Huelga and M. B. Plenio},
  title        = {Noise-assisted energy transfer in quantum networks and light-harvesting complexes},
  journal      = {New Journal of Physics},
  volume       = {12},
  number       = {6},
  pages        = {065002},
  year         = {2010},
  doi          = {10.1088/1367-2630/12/6/065002}
}

@article{Plenio09,
  author       = {F. Caruso and A. W. Chin and A. Datta and S. F. Huelga and M. B. Plenio},
  title        = {Highly efficient energy excitation transfer in light-harvesting complexes: The fundamental role of noise-assisted transport},
  journal      = {The Journal of Chemical Physics},
  volume       = {131},
  number       = {10},
  pages        = {105106},
  year         = {2009},
  doi          = {10.1063/1.3223548}
}

@article{Plenio08,
  author       = {M. B. Plenio and S. F. Huelga},
  title        = {Dephasing-assisted transport: quantum networks and biomolecules},
  journal      = {New Journal of Physics},
  volume       = {10},
  number       = {11},
  pages        = {113019},
  year         = {2008},
  doi          = {10.1088/1367-2630/10/11/113019}
}

@article{Kassal09,
doi = {10.1088/1367-2630/11/3/033003},
url = {https://dx.doi.org/10.1088/1367-2630/11/3/033003},
year = {2009},
month = {mar},
publisher = {},
volume = {11},
number = {3},
pages = {033003},
author = {Rebentrost, Patrick and Mohseni, Masoud and Kassal, Ivan and Lloyd, Seth and Aspuru-Guzik, Alán},
title = {Environment-assisted quantum transport},
journal = {New Journal of Physics}
}

@article{zerah-harush2018,
  title = {Universal Origin for Environment-Assisted Quantum Transport in Exciton Transfer Networks},
  author = {Elinor Zerah-Harush and Yonatan Dubi},
  journal = {The Journal of Physical Chemistry Letters},
  year = {2018},
  volume = {9},
  number = {7},
  pages = {1689-1695},
  doi = {10.1021/acs.jpclett.7b03306}
}

@article{zerah-harush2020,
  title = {Effects of disorder and interactions in environment assisted quantum transport},
  author = {Zerah-Harush, Elinor and Dubi, Yonatan},
  journal = {Phys. Rev. Res.},
  volume = {2},
  issue = {2},
  pages = {023294},
  numpages = {11},
  year = {2020},
  month = {Jun},
  publisher = {American Physical Society},
  doi = {10.1103/PhysRevResearch.2.023294},
  url = {https://link.aps.org/doi/10.1103/PhysRevResearch.2.023294}
}

@article{Kassal2012ordered,
    author = {Kassal, I. and Aspuru-Guzik, A.},
    title = {Environment-assisted quantum transport in ordered systems},
    journal = {New Journal of Physics},
    volume = {14},
    number = {5},
    pages = {053041},
    year = {2012},
    doi = {https://doi.org/10.1088/1367-2630/14/5/053041},
    url = {https://doi.org/10.1088/1367-2630/14/5/053041}
}

@article{kurt2023,
  author       = {Arzu Kurt and Matteo A. C. Rossi and Jyrki Piilo},
  title        = {Quantum transport efficiency in noisy random-removal and small-world networks},
  journal      = {Journal of Physics A: Mathematical and Theoretical},
  volume       = {56},
  number       = {14},
  pages        = {145301},
  year         = {2023},
  doi          = {10.1088/1751-8121/acc0ec}
}

@article{mohseni2014energy,
  title = {Energy-scales convergence for optimal and robust quantum transport in photosynthetic complexes},
  author = {M. Mohseni and A. Shabani and S. Lloyd and H. Rabitz},
  journal = {The Journal of Chemical Physics},
  volume = {140},
  number = {3},
  pages = {035102},
  year = {2014},
  doi = {10.1063/1.4856795}
}

@article{sundelin2026quantumrefrigeration,
  title        = {Quantum refrigeration powered by noise in a superconducting circuit},
  author       = {Sundelin, Simon and Aamir, Mohammed Ali and Kulkarni, Vyom Manish and Castillo-Moreno, Claudia and Gasparinetti, Simone},
  journal      = {Nature Communications},
  volume       = {17},
  number       = {359},
  year         = {2026},
  doi          = {10.1038/s41467-025-67751-z},
  url          = {https://www.nature.com/articles/s41467-025-67751-z},
}

@article{Haenggi90,
  author    = {H{\"a}nggi, P. and Talkner, P. and Borkovec, M.},
  title     = {Reaction-rate theory: fifty years after Kramers},
  journal   = {Rev. Mod. Phys.},
  year      = {1990},
  volume    = {62},
  pages     = {251--341},
  doi       = {10.1103/RevModPhys.62.251}
}

@article{Cao13,
  title = {Generic Mechanism of Optimal Energy Transfer Efficiency: A Scaling Theory of the Mean First-Passage Time in Exciton Systems},
  author = {Wu, Jianlan and Silbey, Robert J. and Cao, Jianshu},
  journal = {Phys. Rev. Lett.},
  volume = {110},
  issue = {20},
  pages = {200402},
  numpages = {5},
  year = {2013},
  month = {May},
  publisher = {American Physical Society},
  doi = {10.1103/PhysRevLett.110.200402},
  url = {https://link.aps.org/doi/10.1103/PhysRevLett.110.200402}
}

@article{Segal00,
  author    = {Segal, Dvira and Nitzan, Abraham and Davis, William B. and Wasielewski, Michael R. and Ratner, Mark A.},
  title     = {Electron Transfer Rates in Bridged Molecular Systems 2. A Steady-State Analysis of Coherent Tunneling and Thermal Transitions},
  journal   = {J. Phys. Chem. B},
  year      = {2000},
  volume    = {104},
  number    = {16},
  pages     = {3817--3829},
  doi       = {10.1021/jp993419y}
}

@article{Pollak89,
  author    = {Pollak, E. and Grabert, H. and H{\"a}nggi, P.},
  title     = {Theory of activated rate processes for arbitrary frequency dependent friction: Solution of the turnover problem},
  journal   = {J. Chem. Phys.},
  year      = {1989},
  volume    = {91},
  number    = {7},
  pages     = {4073--4087},
  doi       = {10.1063/1.457588}
}

@article{Kramer93,
  author    = {Kramer, B. and MacKinnon, A.},
  title     = {Localization: theory and experiment},
  journal   = {Rep. Prog. Phys.},
  year      = {1993},
  volume    = {56},
  number    = {12},
  pages     = {1469--1564},
  doi       = {10.1088/0034-4885/56/12/001}
}

@article{Gluck02,
title = {Wannier–Stark resonances in optical and semiconductor superlattices},
journal = {Physics Reports},
volume = {366},
number = {3},
pages = {103-182},
year = {2002},
issn = {0370-1573},
doi = {https://doi.org/10.1016/S0370-1573(02)00142-4},
url = {https://www.sciencedirect.com/science/article/pii/S0370157302001424},
author = {Markus Glück and Andrey {R. Kolovsky} and Hans Jürgen Korsch}
}

@article{Wannier62,
  title = {Dynamics of Band Electrons in Electric and Magnetic Fields},
  author = {Wannier, Gregory H.},
  journal = {Rev. Mod. Phys.},
  volume = {34},
  issue = {4},
  pages = {645--655},
  numpages = {0},
  year = {1962},
  month = {Oct},
  publisher = {American Physical Society},
  doi = {10.1103/RevModPhys.34.645},
  url = {https://link.aps.org/doi/10.1103/RevModPhys.34.645}
}

@article{Beratan19,
  author    = {Ru, X. and Zhang, P. and Beratan, D. N.},
  title     = {Assessing Possible Mechanisms of Micrometer-Scale Electron Transfer in Heme-Free {Geobacter sulfurreducens} Pili},
  journal   = {J. Phys. Chem. B},
  year      = {2019},
  volume    = {123},
  number    = {24},
  pages     = {5035--5047},
  doi       = {10.1021/acs.jpcb.9b01086}
}

@article{Eshel20,
  author    = {Eshel, Y. and Peskin, U. and Amdursky, N.},
  title     = {Coherence-Assisted Electron Diffusion Across the Multi-Heme Protein-Based Bacterial Nanowire},
  journal   = {Nanotechnology},
  year      = {2020},
  volume    = {31},
  number    = {31},
  pages     = {314002},
  doi       = {10.1088/1361-6528/ab8767}
}

@article {Naggar24,
article_type = {journal},
title = {Long-distance electron transport in multicellular freshwater cable bacteria},
author = {Yang, Tingting and Chavez, Marko S and Niman, Christina M and Xu, Shuai and El-Naggar, Mohamed Y},
editor = {TerAvest, Michaela and Weigel, Detlef},
volume = 12,
year = 2024,
month = {aug},
pub_date = {2024-08-29},
pages = {RP91097},
citation = {eLife 2024;12:RP91097},
doi = {10.7554/eLife.91097},
url = {https://doi.org/10.7554/eLife.91097},
journal = {eLife},
issn = {2050-084X},
publisher = {eLife Sciences Publications, Ltd},
}

@article{NovotnyExp17,
  author    = {Rondin, Loïc and Gieseler, Jan and Ricci, Francesco and Quidant, Romain and Dellago, Christoph and Novotny, Lukas},
  title     = {Direct measurement of Kramers turnover with a levitated nanoparticle},
  journal   = {Nature Nanotechnology},
  year      = {2017},
  volume    = {12},
  pages     = {1130--1133},
  doi       = {https://doi.org/10.1038/nnano.2017.198}
}

@article{kramers1940brownian,
  author  = {Kramers, H. A.},
  title   = {Brownian motion in a field of force and the diffusion model of chemical reactions},
  journal = {Physica},
  year    = {1940},
  volume  = {7},
  number  = {4},
  pages   = {284--304},
  doi     = {10.1016/S0031-8914(40)90098-2}
}

@article{Coates2023,
  author    = {Alexandre R. Coates and Brendon W. Lovett and Erik M. Gauger},
  title     = {From Goldilocks to twin peaks: multiple optimal regimes for quantum transport in disordered networks},
  journal   = {Physical Chemistry Chemical Physics},
  volume    = {25},
  pages     = {10103--10112},
  year      = {2023},
  doi       = {10.1039/D2CP04935J},
  url       = {https://pubs-rsc-org.myaccess.library.utoronto.ca/en/content/articlelanding/2023/cp/d2cp04935j}
}

@article{AA80,
  author  = {Aubry, Serge and Andr{\'e}, Gilles},
  title   = {Analyticity breaking and Anderson localization in incommensurate lattices},
  journal = {Annals of the Israel Physical Society},
  volume  = {3},
  pages   = {133--164},
  year    = {1980}
}

@article{Harper1955,
  author = {Harper, P. G.},
  title = {Single Band Motion of Conduction Electrons in a Uniform Magnetic Field},
  journal = {Proceedings of the Physical Society A},
  volume = {68},
  number = {10},
  pages = {874--878},
  year = {1955},
  doi = {10.1088/0370-1298/68/10/304}
}

@article{Sarma17,
  title = {Mobility edges in one-dimensional bichromatic incommensurate potentials},
  author = {Li, Xiao and Li, Xiaopeng and Das Sarma, S.},
  journal = {Phys. Rev. B},
  volume = {96},
  issue = {8},
  pages = {085119},
  numpages = {18},
  year = {2017},
  month = {Aug},
  publisher = {American Physical Society},
  doi = {10.1103/PhysRevB.96.085119},
  url = {https://link.aps.org/doi/10.1103/PhysRevB.96.085119}
}

@article{Abanin19,
  title = {Colloquium: Many-body localization, thermalization, and entanglement},
  author = {Abanin, Dmitry A. and Altman, Ehud and Bloch, Immanuel and Serbyn, Maksym},
  journal = {Rev. Mod. Phys.},
  volume = {91},
  issue = {2},
  pages = {021001},
  numpages = {26},
  year = {2019},
  month = {May},
  publisher = {American Physical Society},
  doi = {10.1103/RevModPhys.91.021001},
  url = {https://link.aps.org/doi/10.1103/RevModPhys.91.021001}
}

@article{Abanin16,
  author  = {Abanin, Dmitry A. and De Roeck, Wojciech and Huveneers, Fran{\c{c}}ois},
  title   = {Theory of many-body localization in periodically driven systems},
  journal = {Annals of Physics},
  volume  = {372},
  pages   = {1--11},
  year    = {2016},
  doi     = {10.1016/j.aop.2016.05.010}
}

@article{MBLexp,
author = {Jae-yoon Choi  and Sebastian Hild  and Johannes Zeiher  and Peter Schauß  and Antonio Rubio-Abadal  and Tarik Yefsah  and Vedika Khemani  and David A. Huse  and Immanuel Bloch  and Christian Gross },
title = {Exploring the many-body localization transition in two dimensions},
journal = {Science},
volume = {352},
number = {6293},
pages = {1547-1552},
year = {2016},
doi = {10.1126/science.aaf8834},
URL = {https://www.science.org/doi/abs/10.1126/science.aaf8834},
}

@article{Mirlin05,
  title = {Interacting Electrons in Disordered Wires: Anderson Localization and Low-$T$ Transport},
  author = {Gornyi, I. V. and Mirlin, A. D. and Polyakov, D. G.},
  journal = {Phys. Rev. Lett.},
  volume = {95},
  issue = {20},
  pages = {206603},
  numpages = {4},
  year = {2005},
  month = {Nov},
  publisher = {American Physical Society},
  doi = {10.1103/PhysRevLett.95.206603},
  url = {https://link.aps.org/doi/10.1103/PhysRevLett.95.206603}
}

@article{Landi21,
  title = {Dephasing enhanced transport in boundary-driven quasiperiodic chains},
  author = {Lacerda, Artur M. and Goold, John and Landi, Gabriel T.},
  journal = {Phys. Rev. B},
  volume = {104},
  issue = {17},
  pages = {174203},
  numpages = {7},
  year = {2021},
  month = {Nov},
  publisher = {American Physical Society},
  doi = {10.1103/PhysRevB.104.174203},
  url = {https://link.aps.org/doi/10.1103/PhysRevB.104.174203}
}

@ARTICLE{GooldF24,
AUTHOR={Jacob, Samuel L.  and Bettmann, Laetitia P.  and Lacerda, Artur M.  and Zawadzki, Krissia  and Clark, Stephen R.  and Goold, John  and Mendoza-Arenas, Juan José },
TITLE={Dephasing-assisted transport in a tight-binding chain with a linear potential},
JOURNAL={Frontiers in Physics},
VOLUME={12},
YEAR={2024},
URL={https://www.frontiersin.org/journals/physics/articles/10.3389/fphy.2024.1474018},
DOI={10.3389/fphy.2024.1474018}}

@article{WSD23,
  title = {Noise-induced universal diffusive transport in fermionic chains},
  author = {Langlett, Christopher M. and Xu, Shenglong},
  journal = {Phys. Rev. B},
  volume = {108},
  issue = {18},
  pages = {L180303},
  numpages = {6},
  year = {2023},
  month = {Nov},
  publisher = {American Physical Society},
  doi = {10.1103/PhysRevB.108.L180303},
  url = {https://link.aps.org/doi/10.1103/PhysRevB.108.L180303}
}

@article{WSD24,
  title = {Exact dynamics of quantum dissipative XX models: Wannier-Stark localization in the fragmented operator space},
  author = {Teretenkov, Alexander and Lychkovskiy, Oleg},
  journal = {Phys. Rev. B},
  volume = {109},
  issue = {14},
  pages = {L140302},
  numpages = {7},
  year = {2024},
  month = {Apr},
  publisher = {American Physical Society},
  doi = {10.1103/PhysRevB.109.L140302},
  url = {https://link.aps.org/doi/10.1103/PhysRevB.109.L140302}
}

@article{And13,
  author       = {Žnidarič, Marko and Horvat, Matjaž},
  title        = {Transport in a disordered tight-binding chain with dephasing},
  journal      = {The European Physical Journal B},
  year         = {2013},
  volume       = {86},
  number       = {2},
  pages        = {67},
  doi          = {10.1140/epjb/e2012-30730-9},
  url          = {https://doi.org/10.1140/epjb/e2012-30730-9},
  publisher    = {Springer}
}

@article{AnD10,
doi = {10.1088/1367-2630/12/4/043001},
url = {https://doi.org/10.1088/1367-2630/12/4/043001},
year = {2010},
month = {apr},
publisher = {},
volume = {12},
number = {4},
pages = {043001},
author = {Žnidarič, Marko},
title = {Dephasing-induced diffusive transport in the anisotropic Heisenberg model},
journal = {New Journal of Physics}
}

@article{cavity15,
  title = {Observation of Noise-Assisted Transport in an All-Optical Cavity-Based Network},
  author = {Viciani, Silvia and Lima, Manuela and Bellini, Marco and Caruso, Filippo},
  journal = {Phys. Rev. Lett.},
  volume = {115},
  issue = {8},
  pages = {083601},
  numpages = {5},
  year = {2015},
  month = {Aug},
  publisher = {American Physical Society},
  doi = {10.1103/PhysRevLett.115.083601},
  url = {https://link.aps.org/doi/10.1103/PhysRevLett.115.083601}
}

@article{photon24,
  author       = {Tang, Hao and Shang, Xiao-Wen and Shi, Zi-Yu and He, Tian-Shen and Feng, Zhen and Wang, Tian-Yu and Shi, Ruoxi and Wang, Hui-Ming and Tan, Xi and Xu, Xiao-Yun and Wang, Yao and Gao, Jun and Kim, M. S. and Jin, Xian-Min},
  title        = {Simulating photosynthetic energy transport on a photonic network},
  journal      = {npj Quantum Information},
  year         = {2024},
  volume       = {10},
  pages        = {29},
  doi          = {10.1038/s41534-024-00824-x},
  url          = {https://doi.org/10.1038/s41534-024-00824-x},
  publisher    = {Nature Publishing Group},
  note         = {Received: 06 May 2023; Accepted: 19 February 2024; Published: 11 March 2024}
}

@article{TimeD90,
  author       = {Evensky, D. A. and Scalettar, R. T. and Wolynes, P. G.},
  title        = {Localization and Dephasing Effects in a Time-Dependent Anderson Hamiltonian},
  journal      = {The Journal of Physical Chemistry},
  year         = {1990},
  volume       = {94},
  number       = {3},
  pages        = {1149--1154},
  doi          = {10.1021/j100366a027},
  url          = {https://doi.org/10.1021/j100366a027},
  publisher    = {American Chemical Society},
  month        = feb,
  day          = {8}
}

@article{AndR21,
  title = {Subdiffusion in a one-dimensional Anderson insulator with random dephasing: Finite-size scaling, Griffiths effects, and possible implications for many-body localization},
  author = {Taylor, Scott R. and Scardicchio, Antonello},
  journal = {Phys. Rev. B},
  volume = {103},
  issue = {18},
  pages = {184202},
  numpages = {13},
  year = {2021},
  month = {May},
  publisher = {American Physical Society},
  doi = {10.1103/PhysRevB.103.184202},
  url = {https://link.aps.org/doi/10.1103/PhysRevB.103.184202}
}

@article{MBLD16a,
  title = {Robustness of Many-Body Localization in the Presence of Dissipation},
  author = {Levi, Emanuele and Heyl, Markus and Lesanovsky, Igor and Garrahan, Juan P.},
  journal = {Phys. Rev. Lett.},
  volume = {116},
  issue = {23},
  pages = {237203},
  numpages = {5},
  year = {2016},
  month = {Jun},
  publisher = {American Physical Society},
  doi = {10.1103/PhysRevLett.116.237203},
  url = {https://link.aps.org/doi/10.1103/PhysRevLett.116.237203}
}

@article{MBLD16b,
  title = {Dynamics of a Many-Body-Localized System Coupled to a Bath},
  author = {Fischer, Mark H and Maksymenko, Mykola and Altman, Ehud},
  journal = {Phys. Rev. Lett.},
  volume = {116},
  issue = {16},
  pages = {160401},
  numpages = {5},
  year = {2016},
  month = {Apr},
  publisher = {American Physical Society},
  doi = {10.1103/PhysRevLett.116.160401},
  url = {https://link.aps.org/doi/10.1103/PhysRevLett.116.160401}
}

@article{MBLD16c,
  title = {Influence of dephasing on many-body localization},
  author = {Medvedyeva, Mariya V. and Prosen, Toma\ifmmode \check{z}\else \v{z}\fi{} and \ifmmode \check{Z}\else \v{Z}\fi{}nidari\ifmmode \check{c}\else \v{c}\fi{}, Marko},
  journal = {Phys. Rev. B},
  volume = {93},
  issue = {9},
  pages = {094205},
  numpages = {8},
  year = {2016},
  month = {Mar},
  publisher = {American Physical Society},
  doi = {10.1103/PhysRevB.93.094205},
  url = {https://link.aps.org/doi/10.1103/PhysRevB.93.094205}
}

@article{bijay25,
  author       = {Ray, Tamoghna and Ganguly, Katha and Poletti, Dario and Kulkarni, Manas and Agarwalla, Bijay Kumar},
  title        = {Quantum dynamics in lattices in presence of bulk dephasing and a localized source},
  journal      = {arXiv preprint arXiv:2511.00577},
  year         = {2025},
  archivePrefix= {arXiv},
  eprint       = {2511.00577},
  primaryClass = {quant-ph}
}

@article{Scholes14,
    author = {Fassioli, Francesca and Dinshaw, Rayomond and Arpin, Paul C. and Scholes, Gregory D.},
    title = {Photosynthetic light harvesting: excitons and coherence},
    journal = {Journal of The Royal Society Interface},
    volume = {11},
    number = {92},
    pages = {20130901},
    year = {2014},
    month = {03},
    issn = {1742-5689},
    doi = {10.1098/rsif.2013.0901},
    url = {https://doi.org/10.1098/rsif.2013.0901}
}

@article{manzano2016lattice,
    title = {Quantum Transport in d-dimensional lattices},
    author = {Manzano, D. and Chuang, C. and Cao, J.},
    journal = {New Journal of Physics},
    volume = {18},
    number = {4},
    pages = {043044},
    year = {2016},
    doi = {10.1088/1367-2630/18/4/043044},
    url = {https://doi.org/10.1088/1367-2630/18/4/043044}
}

@article{Lahini08andersonphotonic,
  title = {Anderson Localization and Nonlinearity in One-Dimensional Disordered Photonic Lattices},
  author = {Lahini, Yoav and Avidan, Assaf and Pozzi, Francesca and Sorel, Marc and Morandotti, Roberto and Christodoulides, Demetrios N. and Silberberg, Yaron},
  journal = {Phys. Rev. Lett.},
  volume = {100},
  issue = {1},
  pages = {013906},
  numpages = {4},
  year = {2008},
  month = {Jan},
  publisher = {American Physical Society},
  doi = {10.1103/PhysRevLett.100.013906},
  url = {https://link.aps.org/doi/10.1103/PhysRevLett.100.013906}
}

@article{Cao2009,
    title = {Optimization of Exciton Trapping in Energy Transfer Processes},
    author = {Cao, J. and Silbey, R.J.},
    year = {2009},
    journal = {The Journal Of Physical Chemistry A},
    volume = {113},
    issue = {50},
    pages = {13825-13838},
    doi = {10.1021/jp9032589},
    url = {https://doi.org/10.1021/jp9032589}
}

@article{CaoEET1,
    title = {Efficient Energy Transfer in Light-Harvesting Systems, I: optimal temperature, reorganization energy and spatial-temporal correlations},
    author = {Wu, J. and Liu, F. and Shen, Y. and Cao, J. and Silbey, R.J.},
    year = {2010},
    journal = {New Journal of Physics},
    volume = {12},
    issue = {10},
    pages = {105012},
    doi = {10.1088/1367-2630/12/10/105012},
    url = {https://doi.org/10.1088/1367-2630/12/10/105012}
}

@article{CaoEET2,
    title = {Efficient energy transfer in light-harvesting systems: Quantum-classical comparison, flux network, and robustness analysis},
    author = {Wu, Jianlan and Liu, Fan and Ma, JIan and Silbey, Robert J. and Cao, Jianshu},
    year = {2012},
    journal = {Journal of Chemical Physics},
    volume = {137},
    issue = {17},
    pages = {174111},
    doi = {10.1063/1.4762839},
    url = {https://doi.org/10.1063/1.4762839}
}

@article{CaoEET3,
    title = {Efficient Energy Transfer in Light-Harvesting Systems, III: The Influence of the Eight Bacteriochlorophyll on the Dynamics and Efficiency in FMO},
    author = {Moix, J. and Wu, J. and Huo, P. and Coker, D. and Cao, J},
    year = {2011},
    journal = {The Journal Of Physical Chemistry Letters},
    volume = {2},
    issue = {24},
    pages = {3045-3052},
    doi = {10.1021/jz201259v},
    url = {https://doi.org/10.1021/jz201259v}
}

@article{Montiel2014,
    title = {Importance of Excitation and Trapping Conditions in Photosynthetic Environment-Assisted Energy Transport},
    author = {Le\'on-Montiel, R. de J. and Kassal, I. and Torres, J. P.},
    year = {2014},
    journal = {The Journal of Physical Chemistry B},
    volume = {118},
    issue = {36},
    pages = {10588-10594},
    doi = {10.1021/jp505179h},
    url = {https://doi.org/10.1021/jp505179h}
}

@article{biggerstaff2016waveguides,
  title        = {Enhancing coherent transport in a photonic network using controllable decoherence},
  author       = {Biggerstaff, Devon N. and Heilmann, Ren{\'e} and Zecevik, Aidan A. and Gr{\"a}fe, Markus and Broome, Matthew A. and Fedrizzi, Alessandro and Nolte, Stefan and Szameit, Alexander and White, Andrew G. and Kassal, Ivan},
  journal      = {Nature Communications},
  volume       = {7},
  pages        = {11282},
  year         = {2016},
  doi          = {10.1038/ncomms11282},
  url          = {https://doi.org/10.1038/ncomms11282}
}

@article{Shabani2014numerics,
  title = {Numerical evidence for robustness of environment-assisted quantum transport},
  author = {Shabani, A. and Mohseni, M. and Rabitz, H. and Lloyd, S.},
  journal = {Phys. Rev. E},
  volume = {89},
  issue = {4},
  pages = {042706},
  numpages = {7},
  year = {2014},
  month = {Apr},
  publisher = {American Physical Society},
  doi = {10.1103/PhysRevE.89.042706},
  url = {https://link.aps.org/doi/10.1103/PhysRevE.89.042706}
}

@article{Schwartz2007anderson,
    title = {Transport and Anderson localization in disordered two-dimensional photonic lattices},
    author = {Schwartz, T. and Bartal, G. and Fishman, S. and Segev, M.},
    year = {2007},
    journal = {Nature},
    volume = {446},
    issue = {7131},
    pages = {52-55},
    doi = {10.1038/nature05623},
    url = {https://doi.org/10.1038/nature05623}
}

@article{Bandyopadhyay2021WS,
  title = {Dissipative quantum transport in a nanowire},
  author = {Bandyopadhyay, M. and Dattagupta, S.},
  journal = {Phys. Rev. B},
  volume = {104},
  issue = {12},
  pages = {125401},
  numpages = {18},
  year = {2021},
  month = {Sep},
  publisher = {American Physical Society},
  doi = {10.1103/PhysRevB.104.125401},
  url = {https://link.aps.org/doi/10.1103/PhysRevB.104.125401}
}

@incollection{MMSAboxplot,
  author       = {Berk, Kenneth N. and Devore, Jay L. and Carlton, Matthew A.},
  title        = {Overview and Descriptive Statistics},
  booktitle    = {Modern Mathematical Statistics with Applications},
  pages        = {1--48},
  publisher    = {Springer International Publishing},
  year         = {2021},
  doi          = {10.1007/978-3-030-55156-8_1},
  url          = {https://doi.org/10.1007/978-3-030-55156-8_1}
}

@incollection{Spearman,
  author       = {Foreman, David I. and Corder, Gregory W.},
  title        = {Comparing Variables of Ordinal or Dichotomous Scales: Spearman Rank-Order, Point-Biserial, and Biserial Correlations},
  booktitle    = {Nonparametric Statistics},
  publisher    = {John Wiley \& Sons, Incorporated},
  year         = {2014},
  doi          = {10.1002/9781118840429.ch7},
  url          = {https://doi.org/10.1002/9781118165881.ch7},
  pages        = {122-154}
}

@article{Schaller22,
  title = {Nonequilibrium boundary-driven quantum systems: Models, methods, and properties},
  author = {Landi, Gabriel T. and Poletti, Dario and Schaller, Gernot},
  journal = {Rev. Mod. Phys.},
  volume = {94},
  issue = {4},
  pages = {045006},
  numpages = {58},
  year = {2022},
  month = {Dec},
  publisher = {American Physical Society},
  doi = {10.1103/RevModPhys.94.045006},
  url = {https://link.aps.org/doi/10.1103/RevModPhys.94.045006}
}

@article{Erik2020JCP,
  author       = {Davidson, Scott and Fruchtman, Amir and Pollock, Felix A. and Gauger, Erik M.},
  title        = {The dark side of energy transport along excitonic wires: On-site energy barriers facilitate efficient, vibrationally mediated transport through optically dark subspaces},
  journal      = {The Journal of Chemical Physics},
  year         = {2020},
  volume       = {153},
  number       = {13},
  pages        = {134701},
  doi          = {10.1063/5.0023702},
  publisher    = {AIP Publishing},
}

@article{Erik2022PRXQ,
  author       = {Davidson, Scott and Pollock, Felix A. and Gauger, Erik M.},
  title        = {Eliminating Radiative Losses in Long-Range Exciton Transport},
  journal      = {PRX Quantum},
  year         = {2022},
  volume       = {3},
  number       = {2},
  pages        = {020354},
  doi          = {10.1103/PRXQuantum.3.020354},
  publisher    = {American Physical Society}
}

@article{Erik2023PRXE,
  title = {Light Harvesting Enhanced by Quantum Ratchet States},
  author = {Werren, Nicholas and Brown, Will and Gauger, Erik M.},
  journal = {PRX Energy},
  volume = {2},
  issue = {1},
  pages = {013002},
  numpages = {19},
  year = {2023},
  month = {Feb},
  publisher = {American Physical Society},
  doi = {10.1103/PRXEnergy.2.013002},
  url = {https://link.aps.org/doi/10.1103/PRXEnergy.2.013002}
}

@article{GaugerKassal2021JPCL,
  author  = {Tomasi, Silas and Rouse, David M. and Gauger, Erik M. and Lovett, Brendon W. and Kassal, Ivan},
  title   = {Environmentally Improved Coherent Light Harvesting},
  journal = {The Journal of Physical Chemistry Letters},
  year    = {2021},
  volume  = {12},
  number  = {26},
  pages   = {6143--6151},
  doi     = {10.1021/acs.jpclett.1c01303},
}

@article{Coates2021NJP,
  author  = {Coates, Adam R. and Lovett, Brendon W. and Gauger, Erik M.},
  title   = {Localisation Determines the Optimal Noise Rate for Quantum Transport},
  journal = {New Journal of Physics},
  year    = {2021},
  volume  = {23},
  number  = {12},
  pages   = {123014},
  doi     = {10.1088/1367-2630/ac3b2c},
}

@article{Martin2011opticsexpress,
  author       = {Martin, Lane and Di Giuseppe, Giovanni and Perez-Leija, Armando and Keil, Robert and Dreisow, Felix and Heinrich, Matthias and Nolte, Stefan and Szameit, Alexander and Abouraddy, Ayman F. and Christodoulides, Demetrios N. and Saleh, Bahaa E. A.},
  title        = {Anderson localization in optical waveguide arrays with off-diagonal coupling disorder},
  journal      = {Optics Express},
  volume       = {19},
  number       = {14},
  pages        = {13636--13646},
  year         = {2011},
  doi          = {10.1364/OE.19.013636},
}

@article{moore1994PRL,
  title = {Observation of Dynamical Localization in Atomic Momentum Transfer: A New Testing Ground for Quantum Chaos},
  author = {Moore, F. L. and Robinson, J. C. and Bharucha, C. and Williams, P. E. and Raizen, M. G.},
  journal = {Phys. Rev. Lett.},
  volume = {73},
  issue = {22},
  pages = {2974--2977},
  numpages = {0},
  year = {1994},
  month = {Nov},
  publisher = {American Physical Society},
  doi = {10.1103/PhysRevLett.73.2974},
  url = {https://link.aps.org/doi/10.1103/PhysRevLett.73.2974}
}

@article{chabe2008PRL,
  title = {Experimental Observation of the Anderson Metal-Insulator Transition with Atomic Matter Waves},
  author = {Chab\'e, Julien and Lemari\'e, Gabriel and Gr\'emaud, Beno\^{\i}t and Delande, Dominique and Szriftgiser, Pascal and Garreau, Jean Claude},
  journal = {Phys. Rev. Lett.},
  volume = {101},
  issue = {25},
  pages = {255702},
  numpages = {4},
  year = {2008},
  month = {Dec},
  publisher = {American Physical Society},
  doi = {10.1103/PhysRevLett.101.255702},
  url = {https://link.aps.org/doi/10.1103/PhysRevLett.101.255702}
}

@article{Manai2015PRL,
  author       = {Manai, I. and Cl{\'e}ment, J.-F. and Chicireanu, R. and Hainaut, C. and Garreau, J. C. and Szriftgiser, P. and Delande, D.},
  title        = {Experimental Observation of Two-Dimensional Anderson Localization with the Atomic Kicked Rotor},
  journal      = {Physical Review Letters},
  volume       = {115},
  number       = {24},
  pages        = {240603},
  year         = {2015},
  doi          = {10.1103/PhysRevLett.115.240603},
}

@article{Billy2008Anderson,
  author       = {Billy, J. and Josse, V. and Zuo, Z. and Bernard, A. and Hambrecht, B. and Lugan, P. and Cl{\'e}ment, D. and Sanchez-Palencia, L. and Bouyer, P. and Aspect, A.},
  title        = {Direct observation of Anderson localization of matter waves in a controlled disorder},
  journal      = {Nature},
  volume       = {453},
  number       = {7197},
  pages        = {891--894},
  year         = {2008},
  doi          = {10.1038/nature07000},
}

@article{Guo2021WS,
  author       = {Guo, X.-Y. and Ge, Z.-Y. and Li, H. and Wang, Z. and Zhang, Y.-R. and Song, P. and Xiang, Z. and Song, X. and Jin, Y. and Lu, L. and Xu, K. and Zheng, D. and Fan, H.},
  title        = {Observation of Bloch oscillations and Wannier-Stark localization on a superconducting quantum processor},
  journal      = {npj Quantum Information},
  volume       = {7},
  number       = {1},
  pages        = {51},
  year         = {2021},
  doi          = {10.1038/s41534-021-00385-3},
}

@article{Gao2023WS,
  title = {Coexistence of extended and localized states in finite-sized mosaic Wannier-Stark lattices},
  author = {Gao, Jun and Khaymovich, Ivan M. and Iovan, Adrian and Wang, Xiao-Wei and Krishna, Govind and Xu, Ze-Sheng and Tortumlu, Emrah and Balatsky, Alexander V. and Zwiller, Val and Elshaari, Ali W.},
  journal = {Phys. Rev. B},
  volume = {108},
  issue = {14},
  pages = {L140202},
  numpages = {7},
  year = {2023},
  month = {Oct},
  publisher = {American Physical Society},
  doi = {10.1103/PhysRevB.108.L140202},
  url = {https://link.aps.org/doi/10.1103/PhysRevB.108.L140202}
}

@article{Song2024PRXWS,
  title = {Coherent Control of Bloch Oscillations in a Superconducting Circuit},
  author = {Song, Pengtao and Xiang, Zhongcheng and Zhang, Yu-Xiang and Wang, Zhan and Guo, Xueyi and Ruan, Xinhui and Song, Xiaohui and Xu, Kai and Gao, Yvonne Y. and Fan, Heng and Zheng, Dongning},
  journal = {PRX Quantum},
  volume = {5},
  issue = {2},
  pages = {020302},
  numpages = {11},
  year = {2024},
  month = {Apr},
  publisher = {American Physical Society},
  doi = {10.1103/PRXQuantum.5.020302},
  url = {https://link.aps.org/doi/10.1103/PRXQuantum.5.020302}
}

@article{Mukherjee2015,
  author       = {Mukherjee, Sebabrata and Spracklen, Andrew and Choudhury, Debaditya and Goldman, Nathan and {\"O}hberg, Patrik and Andersson, Erika and Thomson, Robert R.},
  title        = {Modulation-assisted tunneling in laser-fabricated photonic Wannier--Stark ladders},
  journal      = {New Journal of Physics},
  year         = {2015},
  volume       = {17},
  number       = {11},
  pages        = {115002},
  doi          = {10.1088/1367-2630/17/11/115002},
}

@article{Creatore2013,
  author    = {C. Creatore and M. A. Parker and S. Emmott and A. W. Chin},
  title     = {Efficient Biologically Inspired Photocell Enhanced by Delocalized Quantum States},
  journal   = {Physical Review Letters},
  volume    = {111},
  pages     = {253601},
  year      = {2013},
  doi       = {10.1103/PhysRevLett.111.253601}
}

@article{Fruchtman2016,
  author    = {Amir Fruchtman and Rafael G{\'o}mez-Bombarelli and Brendon W. Lovett and Erik M. Gauger},
  title     = {Photocell Optimization Using Dark State Protection},
  journal   = {Physical Review Letters},
  volume    = {117},
  number    = {20},
  pages     = {203603},
  year      = {2016},
  doi       = {10.1103/PhysRevLett.117.203603},
  publisher = {American Physical Society}
}

@article{DeSio2017,
  author    = {Antonietta De Sio and Christoph Lienau},
  title     = {Vibronic coupling in organic semiconductors for photovoltaics},
  journal   = {Physical Chemistry Chemical Physics},
  year      = {2017},
  volume    = {19},
  pages     = {18813--18830},
  doi       = {10.1039/c7cp03007j},
  publisher = {The Owner Societies},
  note      = {Published on 26 June 2017},
  url       = {https://pubs.rsc.org/en/content/articlepdf/2017/cp/c7cp03007j}
}

@article{Rouse2019,
  author    = {D. M. Rouse and E. M. Gauger and B. W. Lovett},
  title     = {Optimal power generation using dark states in dimers strongly coupled to their environment},
  journal   = {New Journal of Physics},
  volume    = {21},
  number    = {6},
  pages     = {063025},
  year      = {2019},
  doi       = {10.1088/1367-2630/ab25ca},
  publisher = {IOP Publishing}
}

@article{Cavassilas2020,
  author    = {Nicolas Cavassilas and Daniel Suchet and Amaury Delamarre and Jean-Francois Guillemoles and Fabienne Michelini and Marc Bescond and Michel Lannoo},
  title     = {Optimized Operation of Quantum-Dot Intermediate-Band Solar Cells Deduced from Electronic Transport Modeling},
  journal   = {Physical Review Applied},
  volume    = {13},
  number    = {4},
  pages     = {044035},
  year      = {2020},
  doi       = {10.1103/PhysRevApplied.13.044035},
}

@article{Hu2021QDTransport,
  author    = {Lilei Hu and Andreas Mandelis},
  title     = {Advanced characterization methods of carrier transport in quantum dot photovoltaic solar cells},
  journal   = {Journal of Applied Physics},
  volume    = {129},
  number    = {9},
  year      = {2021},
  doi       = {10.1063/5.0029440},
  url       = {https://doi.org/10.1063/5.0029440}
}

@article{Lei2022ChargeTransportQD,
  author    = {Shiyun Lei and Kanglin Yu and Biao Xiao and Mingrui Zhang and Huan Tao and Liwen Hu and Liyong Zou and Qingliang You and Xunchang Wang and Xueqing Liu and Jiyan Liu and Renqiang Yang},
  title     = {Temperature-dependent transition of charge transport in core/shell structured colloidal quantum dot thin films: From Poole--Frenkel emission to variable-range hopping},
  journal   = {Applied Physics Letters},
  volume    = {121},
  number    = {6},
  year      = {2022},
  doi       = {10.1063/5.0100130},
  url       = {https://doi.org/10.1063/5.0100130}
}

@article{Yang2015QDEfficiency,
  author    = {Yixing Yang and Ying Zheng and Weiran Cao and Alexandre Titov and Jake Hyvonen and Jesse R. Manders and Jiangeng Xue and Paul H. Holloway and Lei Qian},
  title     = {High-efficiency light-emitting devices based on quantum dots with tailored nanostructures},
  journal   = {Nature Photonics},
  volume    = {9},
  pages     = {259--266},
  year      = {2015},
  doi       = {10.1038/nphoton.2015.36},
  url       = {https://www.nature.com/articles/nphoton.2015.36}
}

@article{Bush2021,
  author    = {Robert A. Bush and Erick D. Ochoa and Justin K. Perron},
  title     = {Transport through quantum dots: An introduction via master equation simulations},
  journal   = {American Journal of Physics},
  year      = {2021},
  volume    = {89},
  number    = {3},
  pages     = {300--306},
  doi       = {10.1119/10.0002404},
  publisher = {AIP Publishing},
  url       = {https://doi.org/10.1119/10.0002404}
}

@article{Bahadou2025,
  author    = {M. Ait Bahadou and A. El Allati and K. El Anouz},
  title     = {Quantum transport through a quantum dot coupled to a pair of superconducting topological nanowires},
  journal   = {Physica Scripta},
  volume    = {100},
  number    = {4},
  pages     = {045920},
  year      = {2025},
  doi       = {10.1088/1402-4896/ad9e4a},
  publisher = {IOP Publishing Ltd.}
}

@article{ContrerasPulido2017,
  author    = {L. D. Contreras-Pulido and M. Bruderer},
  title     = {Coherent and incoherent charge transport in linear triple quantum dots},
  journal   = {Journal of Physics: Condensed Matter},
  volume    = {29},
  number    = {18},
  pages     = {185301},
  year      = {2017},
  doi       = {10.1088/1361-648X/aa66d0},
  publisher = {IOP Publishing}
}

@article{Wang2007,
  author    = {Jian-Ming Wang and Rui Wang and Jiu-Qing Liang},
  title     = {Spin-polarized quantum transport through an Aharonov–Bohm quantum-dot-ring},
  journal   = {Chinese Physics},
  volume    = {16},
  number    = {7},
  pages     = {2075},
  year      = {2007},
  doi       = {10.1088/1009-1963/16/7/045},
  publisher = {Chinese Physical Society and IOP Publishing Ltd}
}

@article{Abdullah2016,
  author    = {Nzar Rauf Abdullah and Aziz H. Fatah and Jabar M. A. Fatah},
  title     = {Effects of magnetic field on photon-induced quantum transport in a single dot-cavity system},
  journal   = {Chinese Physics B},
  volume    = {25},
  number    = {11},
  pages     = {114206},
  year      = {2016},
  doi       = {10.1088/1674-1056/25/11/114206},
  publisher = {Chinese Physical Society and IOP Publishing Ltd}
}

@article{Mathe2022,
  author    = {Levente Máthé and Doru Sticlet and Liviu P. Zârbo},
  title     = {Quantum transport through a quantum dot side-coupled to a Majorana bound state pair in the presence of electron-phonon interaction},
  journal   = {Physical Review B},
  volume    = {105},
  number    = {15},
  pages     = {155409},
  year      = {2022},
  doi       = {10.1103/PhysRevB.105.155409},
  publisher = {American Physical Society}
}

@article{Chen2013,
  author    = {Shuguang Chen and Hang Xie and Yu Zhang and Xiaodong Cui and Guanhua Chen},
  title     = {Quantum transport through an array of quantum dots},
  journal   = {Nanoscale},
  volume    = {5},
  pages     = {169--173},
  year      = {2013},
  doi       = {10.1039/C2NR32343E},
  publisher = {Royal Society of Chemistry}
}

@article{Eastham2013,
  title = {Lindblad theory of dynamical decoherence of quantum-dot excitons},
  author = {Eastham, P. R. and Spracklen, A. O. and Keeling, J.},
  journal = {Phys. Rev. B},
  volume = {87},
  issue = {19},
  pages = {195306},
  numpages = {11},
  year = {2013},
  month = {May},
  publisher = {American Physical Society},
  doi = {10.1103/PhysRevB.87.195306},
  url = {https://link.aps.org/doi/10.1103/PhysRevB.87.195306}
}

@article{Tuokkola2025Transmon,
  author       = {Tuokkola, M. and Sunada, Y. and Kivij{\"a}rvi, H. and others},
  title        = {Methods to achieve near-millisecond energy relaxation and dephasing times for a superconducting transmon qubit},
  journal      = {Nature Communications},
  volume       = {16},
  pages        = {5421},
  year         = {2025},
  doi          = {10.1038/s41467-025-61126-0},
}

@article{Dutta2020,
  title = {Long-Range Coherence and Multiple Steady States in a Lossy Qubit Array},
  author = {Dutta, Shovan and Cooper, Nigel R.},
  journal = {Phys. Rev. Lett.},
  volume = {125},
  issue = {24},
  pages = {240404},
  numpages = {7},
  year = {2020},
  month = {Dec},
  publisher = {American Physical Society},
  doi = {10.1103/PhysRevLett.125.240404},
  url = {https://link.aps.org/doi/10.1103/PhysRevLett.125.240404}
}

@article{Plenio17,
  title = {Colloquium: Quantum coherence as a resource},
  author = {Streltsov, Alexander and Adesso, Gerardo and Plenio, Martin B.},
  journal = {Rev. Mod. Phys.},
  volume = {89},
  issue = {4},
  pages = {041003},
  numpages = {34},
  year = {2017},
  month = {Oct},
  publisher = {American Physical Society},
  doi = {10.1103/RevModPhys.89.041003},
  url = {https://link.aps.org/doi/10.1103/RevModPhys.89.041003}
}

@article{Kim2022PRA,
  title = {Relation between quantum coherence and quantum entanglement in quantum measurements},
  author = {Kim, Ho-Joon and Lee, Soojoon},
  journal = {Phys. Rev. A},
  volume = {106},
  issue = {2},
  pages = {022401},
  numpages = {12},
  year = {2022},
  month = {Aug},
  publisher = {American Physical Society},
  doi = {10.1103/PhysRevA.106.022401},
  url = {https://link.aps.org/doi/10.1103/PhysRevA.106.022401}
}

\newpage
\appendix

\section{Additional Results: Steady state coherences}
\label{AppA}

In this Appendix, we provide the structure of steady-state coherences under both uniform and site-optimized dephasing profiles, complementing Fig. \ref{fig:figure4} and Fig. \ref{fig:figure8}, where ratios of elements were displayed.

\subsection{Steady-state coherences under a ramp potential}

\begin{figure*}[htbp]
     \centering
\includegraphics[width=0.7\textwidth] {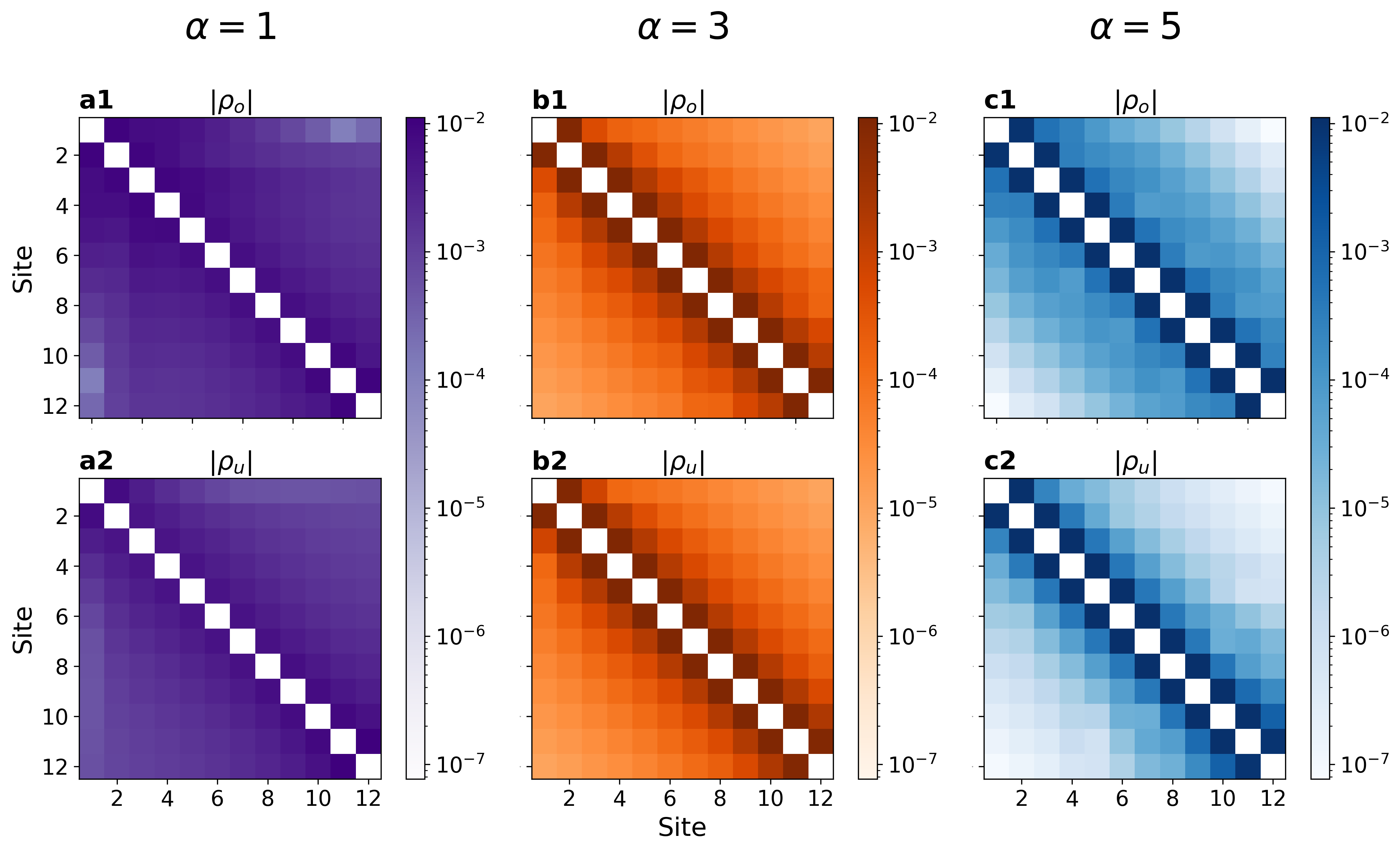}
     \caption{Overcoming WS localization in systems with power-law tunneling with powers of \(\alpha = 1\)  (a1-a2, purple), \(\alpha = 3\) (b1-b2, orange), and \(\alpha = 5\) (c1-c2, blue).
     The energy profile and parameters are presented in Figs. \ref{fig:figure2} and \ref{fig:figure4}. Here we present the steady state solution used to generate the map of ratios in Fig. \ref{fig:figure4} (c), (f), (i).
     Row 1: steady-state density matrices with zero diagonal under the optimized dephasing rates shown in Fig. \ref{fig:figure3}(a).
     Row 2: steady-state density matrices with the optimal uniform dephasing rate, shown by the dashed lines in Fig. \ref{fig:figure3}.
     }
 \label{fig:figure14}
 \end{figure*}

Fig. \ref{fig:figure14} presents the steady-state density matrices (with diagonal elements removed) for systems subject to a ramp potential. Results are shown for three tunneling regimes, $\alpha = 1, 3, 5$, comparing (top) locally optimized dephasing profiles and (bottom) uniform dephasing at the optimal ENAQT rate constant \(\Gamma_\mathrm{u}\).
We make the following observations: First, in the long-range tunneling regime ($\alpha = 1$), both uniform and optimized dephasing lead to extended coherence patterns across the system. 
In contrast, for short-range tunneling ($\alpha = 5$), the coherence structure differs qualitatively. Under uniform dephasing, coherences remain relatively localized near the diagonal. When dephasing is optimized locally, the coherence pattern becomes more extended. 
The intermediate case ($\alpha = 3$) displays no visible extension of coherences under optimization.

\subsection{Steady-state coherences in disordered systems}

\begin{figure*}[htbp]
     \centering
\includegraphics[width=0.7\textwidth]
 {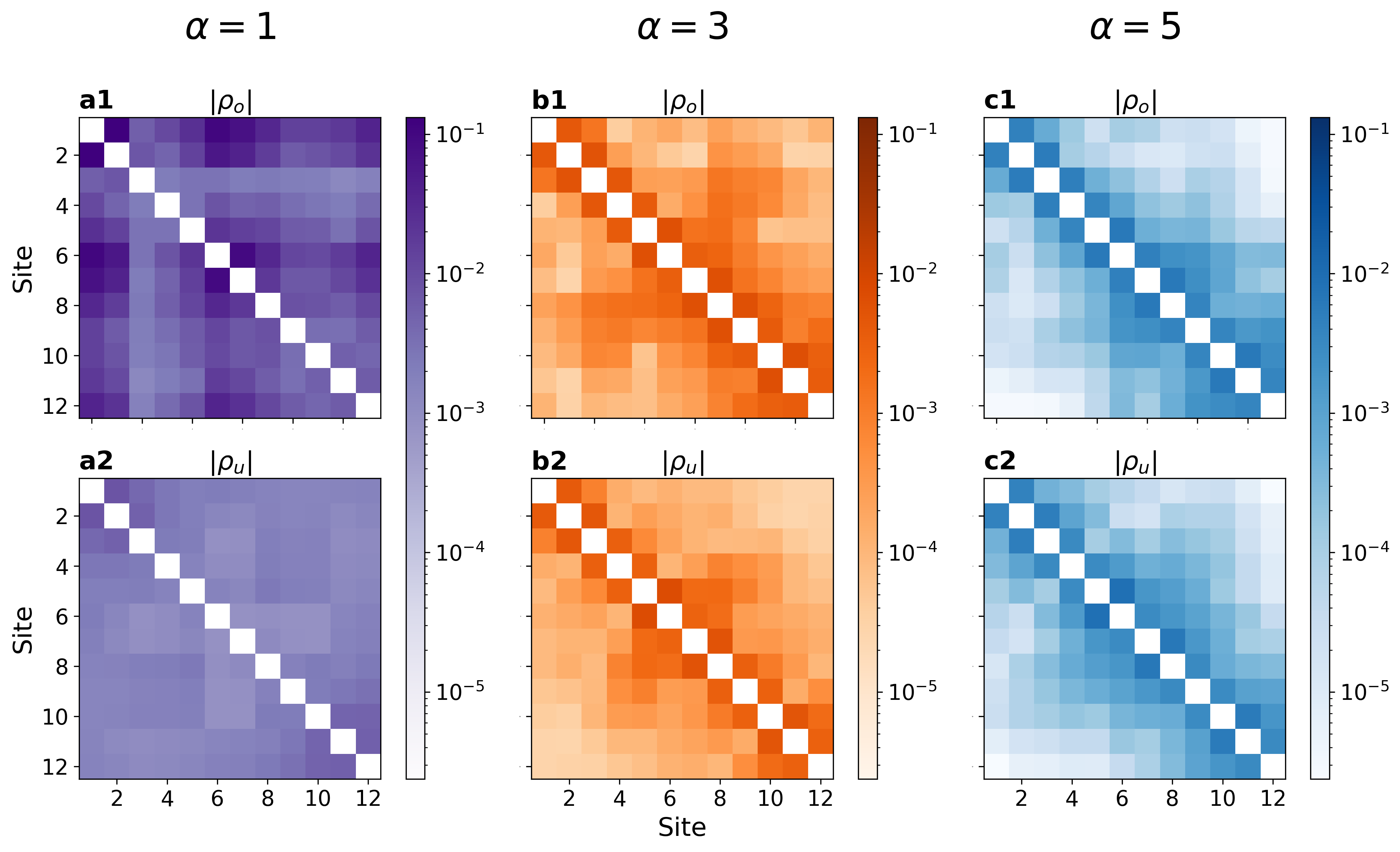}
     \caption{Overcoming Anderson localization in systems with power-law tunneling with powers of \(\alpha = 1\) (a1-a2, purple), \(\alpha = 3\) (b1-b2, orange), and \(\alpha = 5\) (c1-c2, blue).
     Row 1: steady-state density matrices with zero diagonal under the optimized dephasing rates shown in Fig. \ref{fig:figure3}a).
     Row 2: steady-state density matrices with zero diagonal under the optimal uniform dephasing rate, shown by the dashed lines in Fig. \ref{fig:figure7}.
     }
 \label{fig:figure15}
 \end{figure*}

Figure~\ref{fig:figure15} shows maps of state coherences for energy disordered energy landscapes. As in the ramp case, we compare density matrices obtained under site-optimized (top) and uniform dephasing (bottom).

In the long-range tunneling regime ($\alpha = 1$), both uniform and optimized dephasing produce relatively delocalized states, though the optimized case again shows a broader distribution of coherence amplitudes. 
For short-range tunneling ($\alpha = 5$), optimized dephasing leads to a 
nontrivial pattern with more delocalization in the second half of the lattice, away from site 5 that experiences high dephasing. 
The site-optimal dephasing can thus form clusters with stronger delocalization compared to the case with uniform dephasing.




\section{Additional Simulations: Ramp potential with smaller bias}
\label{AppB}

\begin{figure*}[htbp]
    \centering
\includegraphics[width=0.9\linewidth]{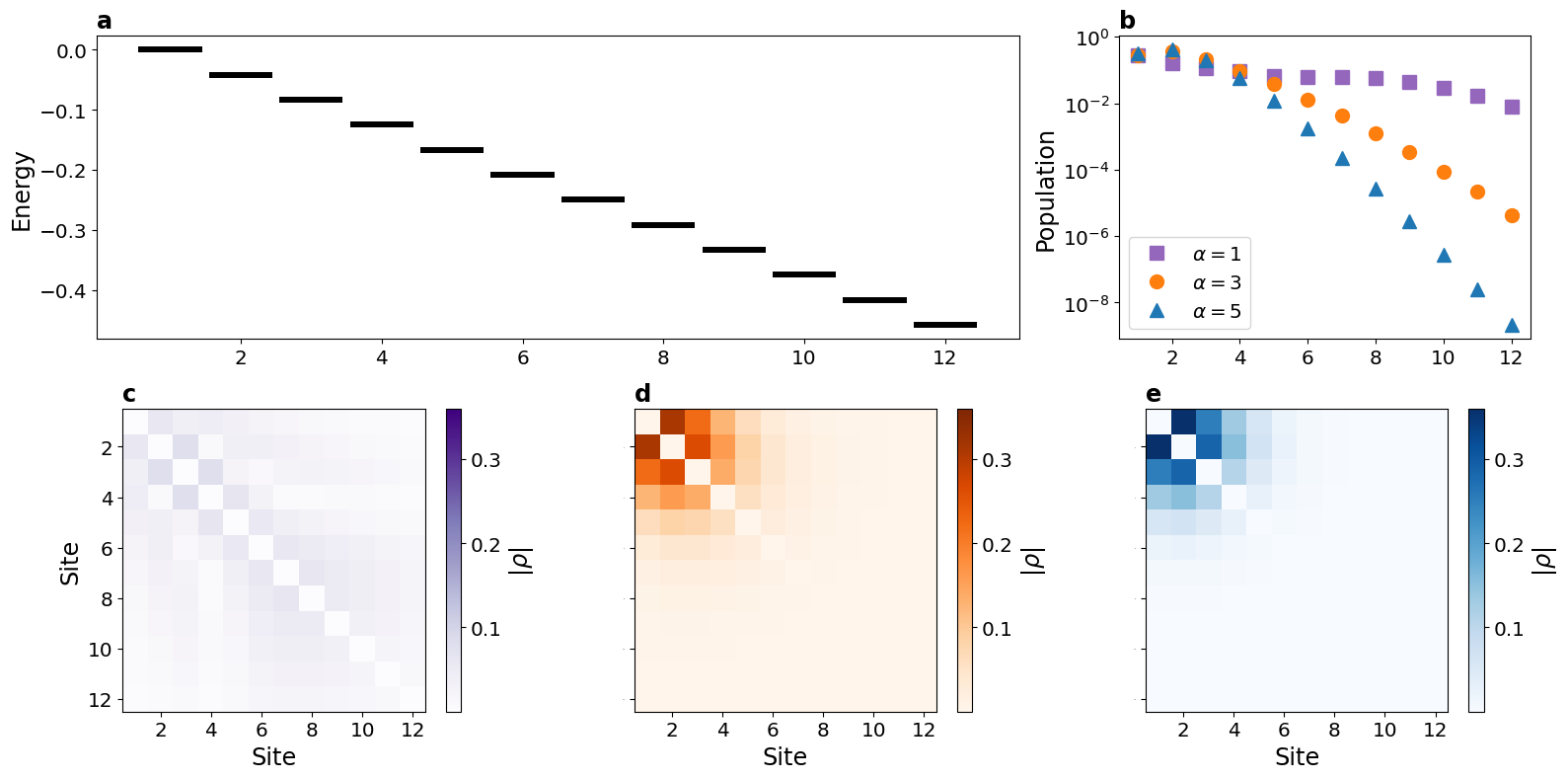}
    \caption{Ramp system: Chains with \(\alpha = 1, 3, 5\) with no environmental interaction. (a) Energy potential profile. The total energy bias is half that of the systems shown in Fig. \ref{fig:figure2}. (b) steady-state population profiles for the ramp energy potential. (c)-(e) steady-state density matrices with the diagonals removed. Other parameters are \(J_{max}\) = 0.1 and \(\gamma_l = 0.1\).}
    \label{fig:figure16}
\end{figure*}

We complement the analysis of the ramp potential Sec. \ref{sec:WS} with another example, this time reducing the potential bias. Fig. \ref{fig:figure16}(a)
presents the ramp potential, with local energy gaps of 0.5/12; the total gap extends from 0 to $-0.46$. Other parameters remain the same as in Fig. \ref{fig:figure2}. First, we examine the system in the absence of dephasing, and show the steady state population profile in Fig. \ref{fig:figure16}(b), and the coherence pattern in Fig. \ref{fig:figure16}(c)-(e), analyzing different tunneling ranges, parameterized by the tunneling power $\alpha$.
Overall, the system does not localize as much as it did at higher bias, given that $J_{max}$ exceeds the nearest neighbor energy spacing $\Delta$. 

\begin{figure}[htbp]
    \centering
\includegraphics[width=0.8\linewidth]{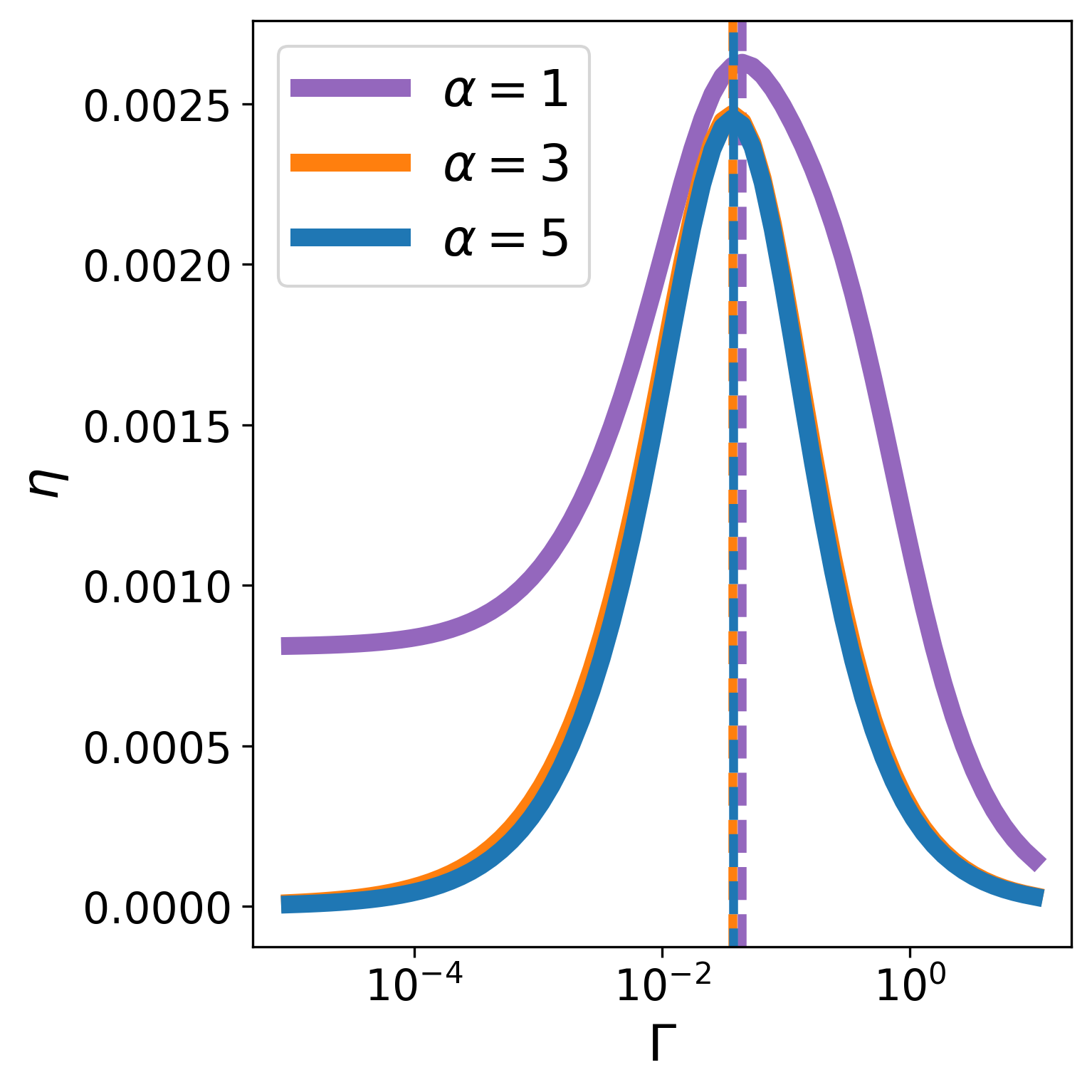}
    \caption{Population flux as a function of uniform dephasing under the ramp potential shown in Fig. \ref{fig:figure16}(a). The peak indicates the optimal dephasing rate constant for transport. The dashed lines mark the locations of the peaks: \(\Gamma_u^{\alpha = 1} = 0.0439\), \(\Gamma_u^{\alpha = 3} = 0.0368\), and \(\Gamma_u^{\alpha = 5} = 0.0374\) with corresponding fluxes $\eta_{\mathrm{u}}^{\alpha = 1}$ = $2.63 \cdot 10^{-3}$, $\eta_{\mathrm{u}}^{\alpha = 3}$ = $2.47 \cdot 10^{-3}$, and $\eta_{\mathrm{u}}^{\alpha = 5}$ = $2.45 \cdot 10^{-3}$. Other parameters are the same as Fig. \ref{fig:figure2}.
    } 
    \label{fig:figure17}
\end{figure}

In Fig. \ref{fig:figure17} we present the population flux under uniform dephasing. We once more observe a standard feature of ENAQT where $\Gamma_u\approx \Delta$; $\Gamma_u$ refers to the dephasing value that optimizes transport under this uniform dephasing. Similarly to Fig. \ref{fig:figure3}, we find that the system conducts better under long range tunneling. 

\begin{figure*}[htbp]
    \centering
\includegraphics[width=1.\linewidth]{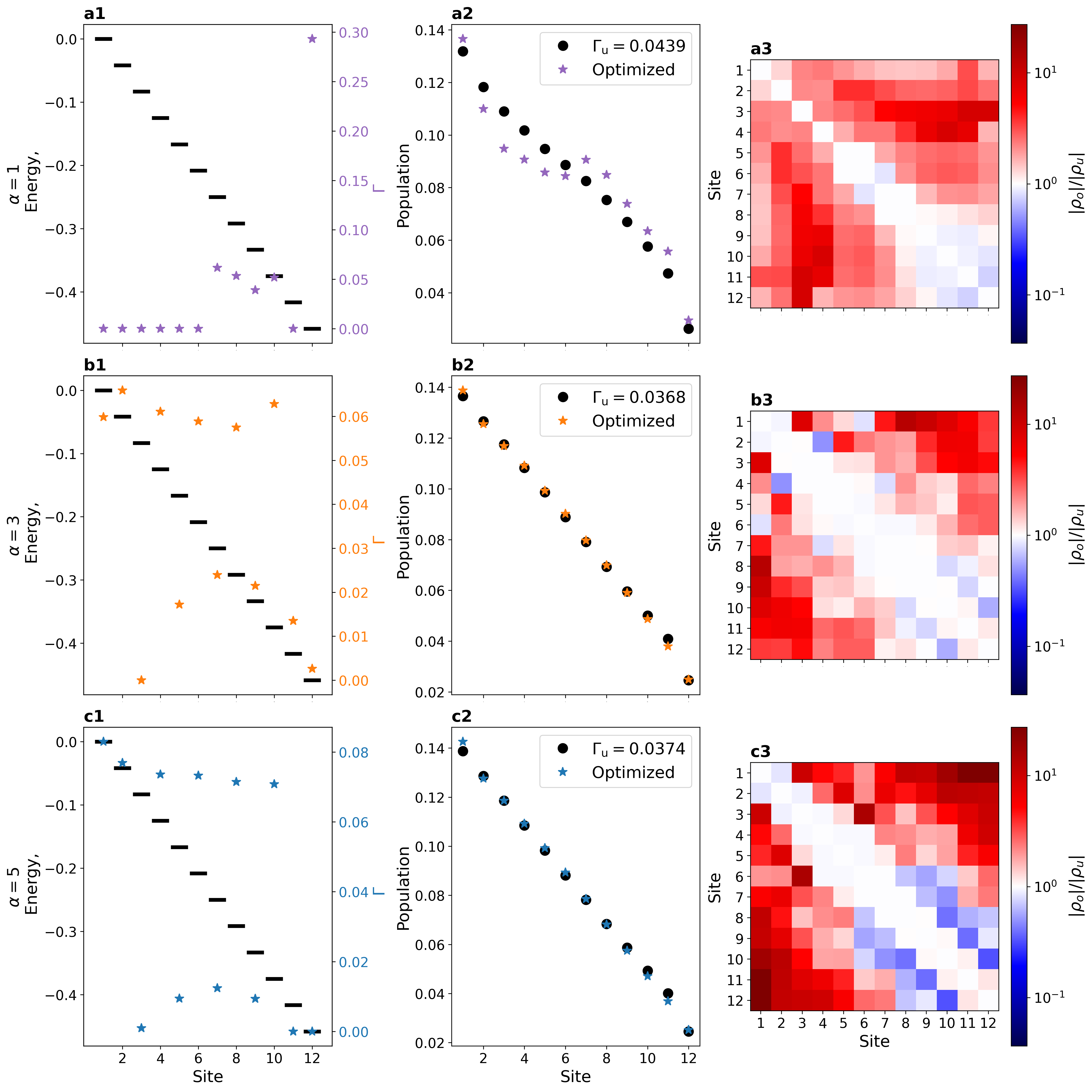}
    \caption{Overcoming WS localization in systems with power-law tunneling (a1-a3) \(\alpha = 1\), (b1-b3) \(\alpha = 3\), (c1-c3) \(\alpha = 5\). Left column: ramp energy profile and the resulting locally optimized values for \(\Gamma_n\). The total energy bias is half that of Fig. \ref{fig:figure3}. Middle column: Steady-state population using \(\Gamma_u\) and under the optimized profile of \(\Gamma_n\). Right column: Ratios of the absolute values of the density matrix elements under the locally optimized dephasing rates \(\rho_o\), and the uniform \(\Gamma_u\) marked in Fig. \ref{fig:figure17}. Other parameters are the same as in Fig. \ref{fig:figure2}.
    }
    \label{fig:figure18}
\end{figure*}

We now optimize the dephasing at each site and present the results compared to the uniform dephasing case (see Fig.~\ref{fig:figure18}). 
For long-range tunneling, and in the presence of a small bias, we find that the first half of the lattice does not require dephasing. Only beyond the midpoint does the system begin to benefit from increasing dephasing. Notably, the final site exhibits strong dephasing, which promotes efficient edge-to-edge transport.
As tunneling becomes more localized ($\alpha = 3,5$), we observe an alternating pattern in the optimized dephasing profile, consistent with the behavior previously identified at larger energy bias, Fig. \ref{fig:figure4}.

Overall, comparing Figs. \ref{fig:figure2}-\ref{fig:figure4}
to Figs. \ref{fig:figure16}-\ref{fig:figure18},
we note that increasing the bias does not change our main observations regarding how dephasing can be optimized to destroy the WS localization and reach larger currents. 

\end{document}